%% file: arxiv.tex
\documentclass[journal,twoside]{IEEEtran}
\IEEEoverridecommandlockouts
\usepackage{cite}

\usepackage{bm}

\ifCLASSINFOpdf
  \usepackage[pdftex]{graphicx}
  \graphicspath{{../images/}}
  \DeclareGraphicsExtensions{.pdf,.jpeg,.png}
\else
  \usepackage[dvips]{graphicx}
  \graphicspath{{../images/}}
  \DeclareGraphicsExtensions{.eps}
\fi
\usepackage[cmex10]{amsmath}
\usepackage{amssymb}
\ifCLASSOPTIONcompsoc
  \usepackage[caption=false,font=normalsize,labelfont=sf,textfont=sf]{subfig}
\else
  \usepackage[caption=false,font=footnotesize]{subfig}
\fi
\usepackage{tikz}

\usepackage{mathrsfs}
\usepackage{multirow}

\usepackage{acronym}

\usepackage{pgfplots} 
\pgfplotsset{compat=newest} 
\pgfplotsset{plot coordinates/math parser=false} 
\newlength\figureheight 
\newlength\figurewidth 
\usepackage{tikzscale}
\usepackage{epstopdf}

\usepackage{cleveref}

\newcommand{\conv}[2]{\left\langle #1 , #2 \right\rangle}

\newcommand{\setC}{\boldsymbol{C}}

\newcommand{\setX}{\boldsymbol{X}}
\newcommand{\setY}{\boldsymbol{Y}}
\newcommand{\numSens}{{N}_{\mathrm{sens}}}
\newcommand{\statef}{\boldsymbol{f}}
\newcommand{\statex}{\boldsymbol{x}}
\newcommand{\statey}{{\boldsymbol{y}}}
\newcommand{\stateyScal}{y}

\newcommand{\setZ}{\boldsymbol{Z}}

\newcommand{\statez}{\boldsymbol{z}}
\newcommand{\diff}{\mathrm{d}}
\newcommand{\assoc}{A}
\newcommand{\assocspace}{\mathcal{A}}

\usepackage{xspace}

\let\oldFootnote\footnote
\newcommand\nextToken\relax
\renewcommand\footnote[1]{%
	\oldFootnote{#1}\futurelet\nextToken\isFootnote}
\newcommand\isFootnote{%
	\ifx\footnote\nextToken\textsuperscript{,}\fi}

\usepackage[normalem]{ulem}
\acrodef{ad}[AD]{autonomous driving}
\acrodef{cdf}[CDF]{cumulative distribution function}
\acrodef{ci}[CI]{covariance intersection}
\acrodef{cphd}[CPHD]{Cardinalized \ac{phd}}
\acrodef{cv}[CV]{constant velocity}
\acrodef{da}[DA]{data association}
\acrodef{dglmb}[$\delta$-GLMB]{$\delta$-generalized labelled multi-Bernoulli}
\acrodef{ekf}[EKF]{extended Kalman filter}
\acrodef{emd}[EMD]{exponential mixture density}
\acrodef{et}[ET]{extended target}
\acrodef{ett}[ETT]{extended target tracking}
\acrodef{etpmb}[PMB-ETT]{\ac{pmb}-\ac{ett}}
\acrodef{fim}[FIM]{Fisher information matrix}
\acrodef{fisst}[FISST]{finite-set statistics}
\acrodef{fg}[FG]{factor graph}
\acrodef{fov}[FoV]{field-of-view}
\acrodef{gg}[G-G]{gamma-Gaussian}
\acrodef{gm}[GM]{Gaussian mixture}
\acrodef{gospa}[GOSPA]{generalized optimal subpattern assignment}
\acrodef{gp}[GP]{Gaussian process}
\acrodef{gnss}[GNSS]{global navigation satellite system}
\acrodef{iou}[IOU]{intersection over union}
\acrodef{jpda}[JPDA]{joint probability data association}
\acrodef{kf}[KF]{Kalman filter}
\acrodef{kla}[KLA]{Kullback-Leibler average}
\acrodef{kld}[KLD]{Kullback-Leibler divergence}
\acrodef{ldm}[LDM]{local dynamic map}
\acrodef{iid}[IID]{independent identically distributed}
\acrodef{imu}[IMU]{inertial measurement unit}
\acrodef{its}[ITS]{intelligent transportation system}
\acrodef{v2v}[V2V]{vehicle-to-vehicle}
\acrodef{v2i}[V2I]{vehicle-to-infrastructure}
\acrodef{lmb}[LMB]{labeled multi-Bernoulli}
\acrodef{mb}[MB]{multi-Bernoulli}
\acrodef{mbm}[MBM]{multi-Bernoulli mixture}
\acrodef{mc}[MC]{Monte-Carlo}
\acrodef{member}[MeMBer]{multiple target multi-Bernoulli}
\acrodef{mtt}[MTT]{multitarget tracking}
\acrodef{mht}[MHT]{multi-hypothesis tracking}
\acrodef{rb}[RB]{Rao-Blackwellization}
\acrodef{rmse}[RMSE]{root mean square error}
\acrodef{rfs}[RFS]{random finite set}
\acrodef{rsu}[RSU]{road side unit}
\acrodef{rv}[RV]{random vector}
\acrodef{slam}[SLAM]{simultaneous localization and mapping}
\acrodef{slat}[SLAT]{simultaneous localization and tracking}
\acrodef{std}[STD]{standard deviation}
\acrodef{tombp}[TOMB/P]{track-oriented marginal MeMBer/Poisson}
\acrodef{pmb}[PMB]{Poisson multi-Bernoulli}
\acrodef{pmbm}[PMBM]{Poisson multi-Bernoulli mixture}
\acrodef{pdf}[PDF]{probability density function}
\acrodef{pf}[PF]{particle filter}
\acrodef{pgfl}[p.g.fl.]{probability generating functional}
\acrodef{ppp}[PPP]{Poisson point process}
\acrodef{ukf}[UKF]{unscented Kalman filter}
\acrodef{v2f}[V2F]{vehicle-to-feature}
\acrodef{ospa}[OSPA]{optimal sub-pattern assignment}
\acrodef{phd}[PHD]{Probability Hypothesis Density}
\acrodef{wrt}[w.r.t.]{with respect to}
\acrodef{v2x}[V2X]{V2X}
\acrodef{gps}[GPS]{Global Positioning System}
\acrodef{rtkgps}[RTK]{Real Time Kinematic}
\acrodef{sbasgps}[SBAS]{Satellite-Based Augmentation System}
\acrodef{spsgps}[SPS]{Standard Positioning System}
\acrodef{dgnss}[DGNSS]{Differential \ac{gnss}}

\begin{document}
\tikzstyle{state}=[shape=circle,draw=blue!90,fill=blue!10,line width=1pt]
%
\title{Decentralized Poisson Multi-Bernoulli Filtering for Vehicle Tracking}

\author{Markus~Fr\"ohle,
Karl~Granstr\"om, 
Henk~Wymeersch 
\thanks{
M. Fr\"ohle was with the Department of Electrical Engineering, Chalmers University of Technology, Gothenburg and is now with Zenuity AB, Sweden. E-mail: \texttt{markus.frohle@zenuity.com}. K. Granstr\"om, and H. Wymeersch are with the Department of Electrical Engineering, Chalmers University of Technology, Gothenburg, Sweden. E-mail: \texttt{ \{karl.granstrom, henkw\}@chalmers.se}.
This work was supported, in part, by the EU-H2020 project HIGHTS (High Precision Positioning for Cooperative ITS Applications) under grant no.~MG-3.5a-2014-636537 and COPPLAR (campus shuttle cooperative perception and planning platform) project funded under grant no. 2015-04849 from Vinnova.
}
}

\maketitle

\begin{abstract}
  A decentralized Poisson multi-Bernoulli filter is proposed to track multiple
  vehicles using multiple high-resolution sensors. Independent filters estimate the
  vehicles' presence, state, and shape using a Gaussian process
  extent model; a decentralized filter is realized through fusion
  of the filters posterior densities. An efficient implementation is
  achieved by parametric state representation, utilization of single
  hypothesis tracks, and fusion of vehicle information based on a
  fusion mapping. Numerical results demonstrate the performance.
\end{abstract}

\acresetall


%
\IEEEpeerreviewmaketitle

%

\section{Introduction}
\input{text/introduction}

\section{Related work}\label{sec:relatedwork}
\input{text/relatedwork}

\section{Background on Random Finite Sets}\label{sec:backgroundRFS}
\input{text/backgroundrfs}
\section{Problem Formulation and System Model}\label{sec:sysModelAndProbFormulation}
\input{text/systemmodel}

\section{Independent PMB-ETT Filter} \label{sec:centralizedPMBFilter}
\input{text/etpmbfilter}

\section{Decentralized Posterior Fusion} \label{sec:decentralizedPMBfiltering}
\input{text/posteriorfusion}
\section{Numerical Results} \label{sec:results}
\input{text/results}
\section{Conclusions}\label{sec:conclusions}
\input{text/conclusions}
\appendices
\input{text/appendix}

\bibliographystyle{IEEEtran}
\bibliography{bib}
%
%
%

\input{text/bios}

\end{document}

%% file: text/introduction.tex

Multitarget tracking (MTT)\acused{mtt}, i.e., tracking of independently moving targets, is important for surveillance and safety applications \cite{bar1995multitarget,mahler2007statistical}. Traditionally, it has been developed for surveillance of the sky using ground-to-air radar sensors. 
An \ac{mtt} filter allows to incorporate the peculiarities of those kind of sensors: false alarm measurements due to clutter; missed detections; unknown measurement-to-target correspondence; and target appearance and disappearance; which are all challenges that arise for radar-like sensors. In many typical \ac{mtt} scenarios, the sensor resolution is low \ac{wrt} the target size, and a reasonable assumption is to model the targets as points having a kinematic state (e.g., position and velocity). An elegant way to track multiple targets is via the \ac{pmbm} filter \cite{GranstromFS:2016_PMBMETT,GranstromFS:2016fusion}, which preserves a \ac{pmbm} form during prediction and update steps.

With the availability of high resolution sensors, the point target assumption does not hold anymore, see, e.g., \cite{granstromETT2017}. For instance, a high resolution Lidar sensor can obtain in one scan multiple detections from a single vehicle \cite{GranstromRMS:2014} or cyclist \cite{GranstromL:2013}. This is because the sensor resolution is high \ac{wrt} the target (here, a vehicle) size. In such an application scenario, the vehicle extent needs to be modeled (and estimated) as well in the \ac{mtt} filter, leading to an \ac{ett} filter. 
In \ac{ett}, \acp{et} give rise to possibly multiple noisy detections, the vehicle extent (shape and size) is a priori unknown and may vary over time, and the objective is to estimate the \ac{et}'s kinematic state as well as its extent \cite{granstromETT2017}. A common model for the \ac{et} extent is based on \ac{gp} modelling \cite{wahlstrom2015extended,hirscher2016multiple}. 

Neither \ac{mtt} nor \ac{ett} filters are bound to single sensors. When multiple sensors are used, it becomes necessary to fuse the information from the different sensors. Centralized fusion consists of transmitting all measurements to a central processing unit. Decentralized fusion consists of local processing (i.e., perform recursive state-space estimation) at each sensor unit, and transmission of the tracking results. Such an approach was proposed in \cite{wang2017distributed}, for fusing two \ac{mb} densities, though it did not account for a \ac{ppp} component.

In this paper, we utilize and extend the shape model from  \cite{wahlstrom2015extended,hirscher2016multiple}, and integrate it into an \ac{ett} filter, which allows tracking of multiple \acp{et}. The resulting \ac{ett} filter's multiobject posterior is of the so-called \ac{pmb} form \cite{williams2015marginal}. Furthermore, we propose a novel fusion strategy that performs fusion separately for the \ac{ppp} and \ac{mb} parts of the \ac{pmb} density.
The implementation of the \ac{pmb} filter yields a tracking filter with low computational cost\footnote{As measured by the average time it takes for one cycle of prediction and update.} locally at each sensor, and globally low computational cost through the introduction of a fusion map based on the \ac{kld} between target tracks. Simulation results demonstrate the performance of the proposed independent \ac{ett} filter as well as of the decentralized \ac{ett} filtering approach.
The main contributions of this paper are:
\begin{itemize}
\item We apply a state-of-the-art \ac{ett} filter  \cite{GranstromFS:2016_PMBMETT,GranstromFS:2016fusion}  and propose a novel distributed fusion strategy based on the \ac{kla} by fusion of the filters' posterior multiobject densities of \ac{pmb} form; and
\item We enable low complexity in distributed fusion through introduction of a fusion map based on \ac{kld} between target tracks.
\item We extend, in Section \ref{sec:ETMeasurementModel}, the \ac{gp} model from \cite{wahlstrom2015extended,hirscher2016multiple} for the \ac{et} shape description to a multisensor scenario incorporating the sensors' state (position, orientation);
\end{itemize}

The remainder of this paper is organized as follows; Section~\ref{sec:relatedwork} presents relevant related work to this paper's work, 
Section~\ref{sec:backgroundRFS} gives some background knowledge on \acp{rfs}, and
Section~\ref{sec:sysModelAndProbFormulation} introduces the system model and the problem formulation. Section~\ref{sec:centralizedPMBFilter} details the proposed \ac{ett} filter, Section~\ref{sec:decentralizedPMBfiltering} presents the decentralized posterior fusion approach using independent \ac{ett} filters.
Simulation results are given in Section~\ref{sec:results}, and conclusions are drawn in Section~\ref{sec:conclusions}.

%% file: text/relatedwork.tex

Here, we discuss related work relevant for this paper's work covering extent modeling for tracking an \ac{et}, \ac{ett}, and information fusion with multiple sensors.

\subsection{Extent modelling}
Different models for the target extent (shape and size) exist in \ac{ett}, which may be classed according to complexity, ranging from assuming a specific geometric shape \cite{koch2008bayesian,GranstromL:2013,GranstromRMS:2014} with, e.g., unknown translation and rotation, to models that describe general shapes, see, e.g., \cite{BaumH:2014,aftab2017gaussian,ozkan2016rao}. Typically, more complex models provide a richer shape description. Two popular models are the random matrix approach \cite{koch2008bayesian}, where the target shape is described by an ellipsoid; and a \ac{gp} based approach \cite{wahlstrom2015extended,hirscher2016multiple}, where a star convex target shape is described by a \ac{gp}. See \cite{granstromETT2017} for an extensive overview of works on \ac{ett}.

\subsection{Tracking multiple extended targets}

For tracking multiple extended targets, several different filters have been developed: \ac{phd} filter  \cite{mahler_FUSION_2009_extTarg,GranstromLO:2012,GranstromO:2012a,SwainC:2012}; \ac{cphd} filter \cite{LundquistGO:2013}; \ac{dglmb} filter \cite{BeardRGVVS:2016}; and \ac{pmbm} filter \cite{GranstromFS:2016_PMBMETT,GranstromFS:2016fusion}. Multiple extended target tracking filters have been applied to tracking of different target types (e.g., for cars \cite{scheel2016using, scheel2017vehicle,michaelis2017heterogeneous}).

The \ac{dglmb} filter and \ac{pmbm} filter are so called multi-object conjugate priors, meaning that if we start with the conjugate density form (\ac{dglmb} or \ac{pmbm}), then all subsequent predicted and updated densities will be of the same form. Based on the conjugate priors, computationally cheaper, approximate filters have been presented. The \ac{lmb} filter is an approximation of the \ac{dglmb} filter, see \cite{BeardRGVVS:2016}. The \ac{pmb} filter is an approximation of the \ac{pmbm} filter, see \cite{williams2015marginal,xia2018extended}. 

The \ac{pmbm} conjugate prior \cite{Williams:2015conjprior} was originally developed for point targets; a \ac{pmbm} conjugate prior for extended targets was presented in \cite{GranstromFS:2016fusion,GranstromFS:2016_PMBMETT}. In several simulation studies it has been shown that, compared to tracking filters built upon labelled \ac{rfs}, the \ac{pmbm} conjugate prior has good performance
for tracking the set of present target states, for both point targets \cite{XiaGSGF:2017,GarciaFernandezWGS:2018,XiaGSGF:2018} and extended targets \cite{GranstromFS:2016fusion,GranstromFS:2016_PMBMETT,XiaGSGFW:2019}. 

The \ac{pmbm} conjugate priors for point targets and extended targets have been shown to be versatile, and have been used with data from Lidars \cite{GranstromRFS:2017,CamentACP:2017,CamentAC:2018,GranstromSRXF:2018}, radars \cite{CamentACP:2017,CamentAC:2018}, and cameras \cite{ScheideggerG:2018,CamentAC:2018}. They have been successfully applied not only to tracking of moving targets, but also mapping of stationary objects \cite{FatemiGSRH:2016_PMBradarmapping}, as well as joint tracking and sensor localisation \cite{FrohleLGW:multisensorPMB}. Thus, it is well motivated to use the \ac{pmbm} filter in this work; specifically, we work with the computationally cheaper \ac{pmb} filter. Developing similar decentralized tracking algorithms for the \ac{dglmb} or \ac{lmb} filter \cite{BeardRGVVS:2016} is a topic for future work.

\subsection{Tracking using multiple sensors}
Different approaches to incorporate measurements from multiple sensors which were acquired within one scan exist, where the simplest may be seen as performing multiple update steps (e.g., Kalman update or likewise) by augmenting the measurement model to incorporate all sensor models, see, e.g., \cite{simon2006optimal}. 

When sensors are geographically separated, measurements from all sensors need to be transmitted to the central processing unit where the filter is run. In the absence of such a central unit or simply due to the limited capacity of the communication channel, one can perform filtering already at the sensor and share only target track information, e.g., the parameters of a known \ac{pdf} family. Through incorporation of target track information obtained by independent filters, a decentralized filtering approach (i.e., information fusion) can be realized without the need for additional prior information. Such methods need to ensure that (unknown) common information obtained by the independent tracking filters is not double counted, e.g., the prior target density \cite{uhlmann1996general}.

Depending on the type of posterior multiobject density in \ac{mtt}/\ac{ett} filtering, several sub-optimal information fusion strategies have been developed based on, e.g, \ac{ci} for Gaussian densities. %
This replaces the product form of Bayes' rule with the \ac{kla}\footnote{In some literature this is known as \ac{emd}.}. \ac{ci} (and consequently \ac{kla}) is a method to fuse information with unknown priors in a robust way in the sense that the fused posterior is conservative and never overconfident about the estimates and thus implicitly sub-optimal \cite{uhlmann1996general,mahler2000optimal}. Examples of \ac{ci}/\ac{kla} include fusion of Bernoulli and \ac{iid} cluster processes posteriors \cite{clark2010robust}, fusion for Bernoulli filters \cite{guldogan2014consensus}, fusion for \ac{phd} and \ac{cphd} filters \cite{uney2010monte, uney2013distributed,li2017generalized, battistelli2013consensus, battistelli2015average, fantacci2015consensus}, and fusion for \ac{lmb} filters \cite{battistelli2015average,wang2017distributed,li2017robust}.

%% file: text/backgroundrfs.tex

Two types of \acp{rfs} relevant for this work deserve special attention: Bernoulli \ac{rfs}, and a \ac{ppp}. They can be extended to \ac{mb} \ac{rfs}, \ac{mbm} \ac{rfs} and combined in a \ac{pmbm} \ac{rfs}. These are described below, for more details on \acp{rfs} the reader is referred to \cite{mahler2014advances}.

\subsection{Common RFS densities}

\subsubsection*{Bernoulli RFS}

A Bernoulli \ac{rfs} $\setX$ has a multiobject density \cite{mahler2014advances}
\begin{align}
f(\setX) = \begin{cases}
1-r, & \setX=\emptyset,\\
r f(\statex), & \setX = \{\statex\},\\
0, & \vert\setX\vert \geq 2,
\end{cases}
\end{align}
where $r\in[0,1]$ denotes the probability that a target exists and if it exists $f(\statex)$ is its \ac{pdf}. 

\subsubsection*{Multi-Bernoulli RFS}
An \ac{mb} \ac{rfs} is the disjoint union of independent Bernoulli \acp{rfs} indexed by $i$. It is fully parametrized by 
$\{r_i,f_i(\statex)\}_{i\in\mathbb{I}}$, where $\mathbb{I}$ is its index set. For $\setX=\{\statex_1,\ldots,\statex_n\}$, the multiobject density can be written as 

\begin{align}
f(\setX) = \sum_{\uplus_{i\in\mathbb{I}}\setX_i=\setX} \prod_{i\in\mathbb{I}}f_{i}(\setX_i)
\end{align}
for $\vert\setX\vert \leq \vert \mathbb{I} \vert$, and $f(\setX) =0$ otherwise. The notation, $\setX_1 \uplus \setX_2 = \setX$ means $\setX_1 \cup \setX_2=\setX$ and $\setX_1 \cap \setX_2=\emptyset$.
\subsubsection*{Multi-Bernoulli Mixture RFS}
The multiobject density of an \ac{mbm} is the normalized, weighted sum of multiobject densities of \acp{mb}, which can be stated as \cite{GranstromFS:2016_PMBMETT}
\begin{align}
f(\setX) = \sum_{j\in\mathbb{J}}w_j\sum_{\uplus_{i\in\mathbb{I}^j}\setX_i=\setX} \prod_{i\in\mathbb{I}^j}f_{j,i}(\setX_i).
\label{eq:definitionMB}
\end{align}
The \ac{mbm} multiobject density is parametrized by $\{w_{j,i},\{r_{j,i},f_{j,i}(\statex)\}_{i\in\mathbb{I}^j}\}_{j\in\mathbb{J}}$, where $w_j$ is the weight of \ac{mb} $j$, and $\mathbb{J}$ is the index set of the \acp{mb} in the \ac{mbm}. An \ac{mb} is therefore a special case of an \ac{mbm} with $|\mathbb{J}|=1$.

\subsubsection*{\ac{ppp}}
A \ac{ppp} is a type of \ac{rfs}, where the cardinality follows a Poisson distribution, and its elements are \ac{iid}. It is parametrized by the intensity function $D(\statex)=\lambda f(\statex)$, where $\lambda>0$ is the Poisson rate and $f(\statex)$ is a \ac{pdf} on the single element state $\statex$. The multiobject density of a \ac{ppp} is \cite{mahler2014advances,GranstromFS:2016_PMBMETT}
\begin{align}
f(\setX)=e^{-\lambda}
\prod_{i=1}^n \lambda f(\statex_i).
\label{eq:definitionPPP}
\end{align}
\subsubsection*{\ac{pmbm}}
An \ac{pmbm} \ac{rfs} is the disjoint set union of an \ac{ppp} and an \ac{mbm} having multiobject density \cite{GranstromFS:2016_PMBMETT}
\begin{align} 
f(\setX) = \sum_{\setX^u \uplus \setX^d = \setX} f^u(\setX^u) %
f^d(\setX^d),
\label{eq:definitionPMBM}
\end{align}
where $f^u(\cdot)$ has multiobject density \eqref{eq:definitionPPP}, and $f^d(\setX^d)$ has multiobject density \eqref{eq:definitionMB}.
For $|\mathbb{J}|=1$, \eqref{eq:definitionPMBM} is called a \ac{pmb} distribution.

\subsection{RFS Bayesian Filter}
\label{sec:bayesianFilterFormulation}
Similar to the \ac{rv} case, an \ac{rfs} based filter can be described, conceptually at least, within the Bayesian filtering framework by performing a prediction step using the motion model \cite[Ch. 14]{mahler2007statistical}
\begin{align}
f_+(\setX)&=\int f(\setX|\setX')f_-(\setX')\delta \setX',
\label{eq:RFSfilterPrediction}
\end{align}
where $f_-(\setX')$ is the prior \ac{rfs} density, $f(\setX|\setX')$
is the multiobject process model,
and a Bayesian update step %
\begin{align}
f(\setX|\boldsymbol{Z})\propto \ell(\boldsymbol{Z}|\setX)f_+(\setX).
\end{align}
Here, $f_+(\setX)$ is the predicted \ac{rfs} density, and $\ell(\boldsymbol{Z}|\setX)$ is the \ac{rfs} measurement likelihood for measurement set $\boldsymbol{Z}$. %

A typical way to estimate the set states from a Bernoulli process with \ac{rfs} density $f(\setX)$ is by comparing the probability
of existence $r$ to an existence threshold $r_{\mathrm{th}}$. For $r>r_{\mathrm{th}},$ the target is
said to exist and has \ac{pdf} $f(\statex)$. Its state can then be estimated by the mean $\hat{\statex} = \int \statex f(\statex)\diff \statex$. See, e.g., \cite{GarciaFernandezWGS:2018} for a elongated discussion on multiobject estimation.

%% file: text/systemmodel.tex

Here, we present first the problem formulation of this paper followed by the \ac{et} state and transition model and the \ac{et} set measurement likelihood function.

\subsection{Problem Formulation}
We consider a scenario with $\numSens$ sensing systems (each composed of sensors plus filter), each collecting measurements using its local sensors, with the aim to jointly surveil an environment $\mathcal{E}$ where vehicles pass (c.f. Fig.~\ref{fig:scenarioIntro}).  
Our goals are (i) to derive a low-complexity \ac{pmb}-\ac{ett} filter for each sensing system $s= 1,2,\ldots,\numSens$, computing in every time step $k$ the posterior density $f_s(\setX_k|\setZ_{s,k:1})$ of the \acp{et}, using only its own sensors with measurement set $\setZ_{s,k:1}$%
; and (ii) to derive a decentralized method to combine posterior information of \acp{et} obtained by $\numSens$ independent \ac{ett} filters in order to obtain a global posterior density $\bar{f}_w(\setX_k|\setZ_{1,k:1},\setZ_{2,k:1},\ldots,\setZ_{\numSens,k:1})$ (i.e., using only the \emph{posterior} densities of each \ac{ett} filter).%

\begingroup
\newlength{\xfigwd}
\setlength{\xfigwd}{\columnwidth}
\begin{figure} 
	\centering
	\footnotesize
	\includegraphics[width=0.9\linewidth]{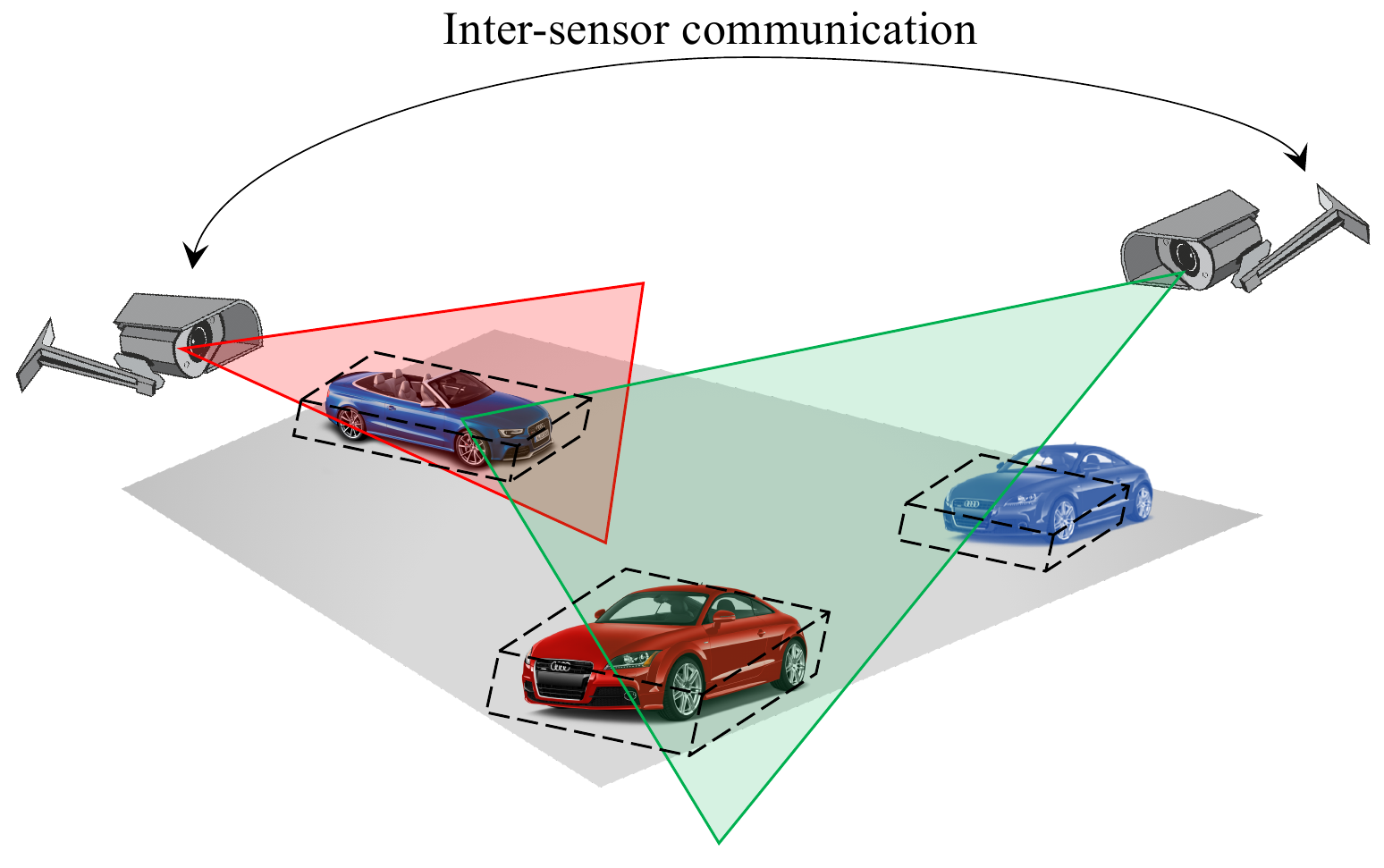}
	\caption{Scenario with three \acp{et} observed by two sensors with partially overlapping \acp{fov}. 
		\label{fig:scenarioIntro}}
\end{figure}
\endgroup

\subsection{ET State and Transition Model}\label{sec:ETstatetransitionmodel}

A standard motion model for the \ac{et} is assumed, where \ac{et} motion follows \ac{iid} Markov processes with single \ac{et} transition \ac{pdf} $f_{k+1|k}(\statex_{k+1}|\statex_k)$, where $\statex_k$ denotes the state at time $k$. Targets arrive according to a non-homogeneous \ac{ppp} with intensity $D^b(\statex_k)$, and depart according to \ac{iid} Markov processes, where the survival probability in $\statex_k$ is $p_\mathrm{S}(\statex_k)$. 
Additionally to this, $\statex_k$ comprises an unknown Poisson rate $\gamma_k$ describing the average number of measurements generated by the \ac{et} and the \ac{et} spatial state $\statey_k$ (this will become clear in Sec.~\ref{sec:ETmodel} where a detailed description of the state and transition density is provided). 

\subsection{ET Set Measurement Likelihood Function}

In one scan, a sensor receives a \emph{set of measurements} $\setZ$ consisting of target-generated measurements $\boldsymbol{z}$ and clutter, 
where the \acp{et} are independently detected with state-dependent probability of detection $p_\mathrm{D}(\statex)$, which depends on the sensor \ac{fov}.  
Clutter is modeled by a \ac{ppp} with intensity $\kappa(\statez)=\lambda c(\statez)$ with mean $\lambda$ and spatial distribution $c(\statez)$. Target-generated measurements are modeled by a \ac{ppp} with intensity $\gamma(\statex)f(\statez|\statex)$, where both the Poisson measurement rate $\gamma(\statex)$ and the single measurement likelihood $f(\statez|\statex)$ are state dependent. The measurement likelihood for \acp{et} $\{\statex_1,\ldots,\statex_n\}$ and measurement set $\setZ$ is
\begin{align}
& \ell( \setZ | \{\statex_1,\ldots,\statex_n\})= \nonumber\\ 
&  e^{-\lambda} \sum_{ \setZ_c \uplus \setZ_1 \uplus \ldots \setZ_n = \setZ} [c(\cdot)]^{\setZ_c}\prod_{i=1}^{n}\ell_{\setZ_i}  (\statex_i),\label{eq:fullLikelihood}
\end{align}
where $[c(\cdot)]^{\setZ_c}$ is shorthand for $\prod_{\statez \in \setZ_c} c(\statez)$, $[c(\cdot)]^\emptyset=1$ by definition, and \cite{GranstromFS:2016_PMBMETT}
\begin{align}
\ell_{\setZ} (\statex) = \begin{cases} p_{\mathrm{D}}(\statex)e^{-\gamma(\statex)}
\prod_{\statez \in \setZ} \gamma(\statex)f(\statez|\statex), & |\setZ|>0,\\
(1-p_\mathrm{D}(\statex))+p_\mathrm{D}(\statex)e^{-\gamma(\statex)}, & \setZ=\emptyset.
\end{cases}
\label{eq:ETTmeasurementModel}
\end{align}
Note that (i) equation \eqref{eq:fullLikelihood} involves potentially multiple \acp{et}, leading to a \ac{da} problem; (ii) a single \ac{et} can generate multiple measurements. 

%% file: text/etpmbfilter.tex

In this section, we briefly describe the processing performed by each \ac{etpmb} filter from \cite{GranstromFS:2016_PMBMETT}. We omit the time index $k$ for brevity. 
The \ac{pmb} model is a combination of a \ac{ppp} describing the distribution of unknown targets, i.e., targets which are hypothesized to exist, but have not yet been detected; and a \ac{mb} which describes targets that have been detected at least once. The target set can therefore be split into two disjoint subsets $\setX = \setX^u \uplus \setX^d$
corresponding to the unknown target set $\setX^u$ (with \ac{ppp} intensity $D^u(\statex)$, {modelled as a non-normalized mixture density with mixture components located in the region of interest, denoted $\mathcal{E}$}) and the detected target set $\setX^d$ with density $f^d(\setX^d)$ and index set $\mathbb{I}$. Hence, the \ac{pmb} density is fully described by a \ac{mb} component described by $\{r^i, f^i(\statex)\}_{i \in \mathbb{I}}$ and a \ac{ppp} component $D^u(\statex)$. 
\acused{etpmb} 

\subsubsection{\ac{etpmb} Filter Prediction}
The predicted density is a \ac{pmb} density with parameters \cite[Sec.~IV]{GranstromFS:2016_PMBMETT}, \cite{GranstromFS:2016fusion}
\begin{align}
D_+^u =& D^b(\statex)+\conv{D^u}{p_\mathrm{S} f_{k+1|k}},\label{eq:predictedPPP}\\
r_+^i =& \conv{f^i}{p_\mathrm{S}}r^i,\label{eq:predictedprobExist}\\
f_+^i =& \frac{\conv{f^i}{p_\mathrm{S} f_{k+1|k}}}{\conv{f^i}{p_\mathrm{S}}},\label{eq:predictedPdf}
\end{align}
where $\conv{g}{h} = \int g(\statex) f(\statex)\diff\statex$ denotes the inner product. 
The proof of the prediction step can be found in, e.g., \cite{williams2015marginal}.

\subsubsection{\ac{etpmb} Filter Update} \label{sec:etpmbFilterUpdateStep}
We introduce the set of valid \acp{da} $\assocspace=\mathcal{P}(\mathbb{M}\cup\mathbb{I})$, where $\mathbb{M}$ is the index set for $\setZ$. Here, $A \in \assocspace$ is a partition of $\mathbb{M}\cup\mathbb{I}$ into non-empty disjoint subsets $C \in A$ (called index cells), with the constraint that for each $C: |C \cap \mathbb{I}| \le 1$ (i.e., measurements can only be associated with a single target). When $|C \cap \mathbb{I}| =1$, let the entry in $C \cap \mathbb{I}$ be denoted by $i_C$ and let $\setC_C = \cup_{m \in C \cap \mathbb{M}}\statez_m$ contain the associated measurements.  Given the predicted prior \ac{pmb} density with parameters \eqref{eq:predictedPPP}, \eqref{eq:predictedprobExist}, \eqref{eq:predictedPdf}, and a set of measurements $\setZ$; the updated density is a \ac{pmbm} density \cite[Sec.~IV]{GranstromFS:2016_PMBMETT}, \cite{GranstromFS:2016fusion}
\begin{align}
f(\setX|\setZ) =& \sum_{\setX^u \uplus \setX^d = \setX} f^u(\setX^u)\sum_{\assoc\in\assocspace}
w_\assoc f_\assoc^d(\setX^d),\label{eq:pmbmposterior}\\
f^u(\setX^u) =& e^{-\conv{D^u}{1}} \prod_{\statex \in \setX^u} D^u(\statex),\label{eq:posteriorPPP}\\
f_\assoc^d(\setX^d) =& \sum_{\uplus_{C\in\assoc} \setX^C=\setX^d} \prod_{C\in\assoc}f_C(\setX^C), \label{eq:PMBFilterfd}\\
D^u(\statex)=& q_\mathrm{D}(\statex)D_+^u(\statex),
\end{align}
where $f_C(\setX^C)$ is a Bernoulli density, with existence probability and spatial distribution provided in Appendix \ref{sec:Appendixetpmb}, together with the   weights $w_\assoc$ of each \ac{da} hypothesis. Above, $q_\mathrm{D}(\boldsymbol{x})$ denotes the probability that the target $\boldsymbol{x}$ is not detected and is defined as
\begin{align}
q_\mathrm{D}(\boldsymbol{x}) = 1-p_\mathrm{D}(\boldsymbol{x}) + p_\mathrm{D}(\boldsymbol{x})e^{-\gamma(\boldsymbol{x})}.
\end{align}
To reduce computational complexity, we use standard methods and truncate the space of possible partitions by clustering measurements and consider \ac{da} \ac{wrt} different clusters \cite{GranstromLO:2012,GranstromO:2012a,LundquistGO:2013}. 
Finally, the \ac{pmbm} in \eqref{eq:pmbmposterior} is converted to a PMB, which was detailed in \cite{Williams:2015}  for point targets and in \cite{xia2018extended} for \acp{et}.

%% file: text/posteriorfusion.tex

Here, we present the decentralized approach to robust fusion of posterior densities $f_s(\setX|\setZ_s)$ computed by independent \ac{etpmb} filters $s$ with unknown prior densities.  
Fusion can be performed after every update step of the filters or based on a lower rate depending on the application and communication capabilities. 
\subsection{Robust Posterior Fusion: Principle} \label{sec:Robustfusion}
Robust posterior fusion can be achieved by minimizing the \ac{kla} between  \ac{rfs} densities $f(\cdot)$ and $f_s(\cdot)$ for $s=1,\ldots,\numSens$
with respect to $f(\cdot)$.  The \ac{kla} is defined as \cite{battistelli2013consensus}
\begin{align}
\bar{f}_{\omega} = \arg \inf_f \sum_{s=1}^{{\numSens}} \omega_s D(f\Vert f_s),
\label{eq:fKLA}
\end{align}
for any combinations of  weights $\omega_s \in [0,1]: \sum_s \omega_s =1$. We have introduced $D(f\Vert f_s)$ as the \ac{kld} between \ac{rfs} densities $f(\setX|\setZ_{1:\numSens})$ and $f_s(\setX|\setZ_s)$ defined as \cite{battistelli2013consensus}
\begin{align}
D(f\Vert f_s) = \int f(\setX|\setZ_{1:\numSens})\log\frac{f(\setX|\setZ_{1:\numSens})}{f_s(\setX|\setZ_s)}\delta\setX.
\label{eq:definitionKLD}
\end{align}
The fused posterior \eqref{eq:fKLA} is robust in the sense that it is conservative and never overconfident \ac{wrt} the true target uncertainty \cite{uhlmann1996general}. Problem \eqref{eq:fKLA} was shown to have closed-form solution \cite{battistelli2013consensus}
\begin{align}
\bar{f}_{\omega}(\setX|\setZ_{1:\numSens})=
\frac{\prod_{s=1}^{\numSens} {f}_s(\setX|\setZ_s)^{{\omega}_s}}{\int \prod_{s^\prime=1}^{\numSens}{f}_{s^\prime}(\setX|\setZ_{s^\prime})^{{\omega}_{s^\prime}} \delta \setX}.
\label{eq:fusedPosteriorMahler}
\end{align}
Note that \eqref{eq:fusedPosteriorMahler} is a generalization of the Uhlmann-Julier covariance intersection method (c.f.~\cite{uhlmann1996general}) for posterior \ac{rfs} densities with unknown priors \cite{mahler2000optimal}. %
The weights $\omega_i$ can be chosen such that \eqref{eq:fusedPosteriorMahler} is as peaky as possible \cite{mahler2000optimal}.

\subsection{Robust PMB Posterior Fusion} \label{sec:PMBfusion}

The posterior density of each \ac{etpmb} filter is a \ac{pmb}, therefore, to be able to utilize the fused posterior density as prior for the next time step in each \ac{etpmb} filter, the fused posterior should be of the \ac{pmb} form, i.e.,
\begin{align}
\bar{f}_{\omega}(\setX|\setZ_{1:\numSens})=&\sum_{\setX^u \uplus \setX^d = \setX}\bar{f}_{\omega}^u(\setX^u|\setZ_{1:\numSens})\nonumber\\
&\times \bar{f}_{\omega}^d(\setX^d|\setZ_{1:\numSens})
\end{align} 
comprised of an \ac{ppp} $\bar{f}_{\omega}^u(\setX^u|\setZ_{1:\numSens})$ modeling the unknown targets and an \ac{mb} $\bar{f}_{\omega}^d(\setX^d|\setZ_{1:\numSens})$ modeling detected targets. {Note, for brevity we avoid the conditioning on the measurement set in the remainder of this section.}

{
A challenge in the fusion step is the fact that sensors do not have the same \ac{fov}s (c.f. Fig.~\ref{fig:scenarioIntro}). In the process of fusion, any target that is in the \ac{fov} of one sensor and has been detected, but outside the \ac{fov} of another sensor, must be treated carefully. In this situation, prior to the fusion, for the first sensor the target corresponds to a detected target represented by a Bernoulli density, whereas for the second sensor the target corresponds to an unknown target, which is represented by the \ac{ppp}. Because of this, in the closed form \ac{kla} solution \eqref{eq:fusedPosteriorMahler}, we must fuse a Bernoulli with a part of the \ac{ppp} intensity. From this it follows that if we solve the \ac{kla} \eqref{eq:fKLA} separately for the \ac{ppp} and the \ac{mb} then we will obtain incorrect results. To enable a valid fusion, we propose the approach outlined in the following sub-sections. 
}

{
Consider \ac{pmb} densities with \ac{ppp} intensity $\lambda_s(x)$ and $N_s$ Bernoullis with parameters $r_{s}^{i}$ and $f_{s}^{i}(x)$ indexed $i=1,\ldots,N_s$. Fusion of the \ac{pmb} densities is simplified if all have the same number of Bernoullis, however, in the general case one cannot assume this. Instead we rely on a result from \cite[Sec. 4.3.1]{mahler2014advances}: any PPP with intensity $\lambda_s(x)$ can be divided into multiple independent PPPs with intensities $\lambda_s^{j}(x)$, where $\sum_{j} \lambda_s^{j}(x) = \lambda_s(x)$. Based on this result, for each sensor we divide $\lambda_s(x)$ into $M_s$ parts such that $N_s+M_s = K$ for all sensors $s$, {where parameter $K$ is determined depending on the number of components of the \acp{mb}.} %
{A robust choice is $K=\sum_{s}N_s$, so that each Bernoulli component in one sensor can be assigned to any combination of PPP or Bernoullis in the other sensors. The \ac{fov} can be taken into account in order to reduce $K$.\footnote{In particular, we partition the deployment region into $2^{\numSens}$ sub-regions, each determined by a subset of sensors that have each partition in the \ac{fov}. For each sub-region $l\in \{1,\ldots,2^{\numSens}\}$, sensor $s$ has $N_{s,l}$ Bernoulli components (with $N_s=\sum_l N_{s,l}$), we determine $K_l=\max N_{s,l}$ and $K=\sum_l K_l$. 
}}

If follows from the division into $K$ parts that the \ac{pmb} densities can be expressed as follows,
\begin{align}
    f_s(\setX) & = \sum_{\uplus_{i=1}^{K}\setX^i = \setX } \prod_{i=1}^{K} f_{s}^{i}(\setX^i)
    \label{eq:MultiP_MultiB_density}
\end{align}
where $f_{s}^{i}(\setX^i)$ is Bernoulli for $i\in\left\{1,\ldots,N_s\right\}$ and \ac{ppp} with intensity $\lambda_{s}^{i}(x)$ for $i\in\left\{N_s+1,\ldots,N_s+M_s\right\}$. Note that the sum in \eqref{eq:MultiP_MultiB_density} is implicit and never has to be computed; it is sufficient to represent the parameters of the densities $f_{s}^{i}(\setX^i)$.

We are seeking the fused density $\bar{f}(\setX)$ that minimizes the KLA \eqref{eq:fKLA}. 
Assume that $\bar{f}(\setX)$ is of the format \eqref{eq:MultiP_MultiB_density}, i.e.,
\begin{align}
    \bar{f}(\setX) & = \sum_{\uplus_{i=1}^{K}\setX^i = \setX } \prod_{i=1}^{K} \bar{f}^{i}(\setX^i).
\label{eq:fBarX}
\end{align}
In \cite{Williams:2015} it is shown that, under this assumption, the minimisation problem \eqref{eq:fKLA} can be solved approximately by minimizing an upper bound,
\begin{align}
    \sum_{s=1}^{\numSens}
    \omega_s \left[ \sum_{i=1}^{K} D\left( \bar{f}^{i}(\setX) || f_{s}^{\pi_s(i)}(\setX) \right) \right]
    \label{eq:KLAupperBound}
\end{align}
where $\pi_s\in\Pi_{(1:K)}$ for all $s$, and $\Pi_{(1:K)}$ is the set of all permutations of the integers $\left\{1,\ldots,K\right\}$. For $\pi\in\Pi$ and $i\in\{1,\ldots,K\}$ we have $\pi(i)\in\left\{1,\ldots,K\right\}$ and $\pi(i)\neq\pi(i')$ for $i\neq i'$. The fusion results will depend on how the permutations $\pi_s$ are chosen, which we discuss in Section~\ref{sec:bestFusionMap}.

Based on \eqref{eq:KLAupperBound}, we compute the $K$ components of the fused density $\bar{f}(\setX)$ as
\begin{align}
    \bar{f}^{i}(\setX) & = \frac{ \prod_{s=1}^{\numSens} \left(f_{s}^{\pi_{s}(i)}(\setX)\right)^{\omega_s} }{ \int \prod_{s=1}^{\numSens} \left(f_{s}^{\pi_{s}(i)}(\setX)\right)^{\omega_s} \delta \setX }.
\end{align}

In summary, our proposed approach to finding the (approximately) optimal fused PMB density consists of the following steps:
\begin{enumerate}
    \item Depending on how the \acp{fov} overlap, we select how to divide the PPP intensities such that the PMB densities all have $K$ parts, and we find permutations $\pi_s$ such that the fused densities $f_{s}^{\pi_{s}(i)}(\setX)$ have a high degree of similarity in the sense of the KL-divergence. Note that this entails determining $K$, $\lambda_{s}^{j}(x)$, and $\pi_{s}$.
    \item Fuse the matched PPPs and Bernoullis, see details below in Section~\ref{sec:BernoulliPPPfusion}.
    \item After the fusion of the $K$ parts of $\bar{f}(\setX)$, we add all the PPP intensities such that a single PPP is obtained.
    \item Lastly, we recycle any Bernoullis in $\bar{f}(\setX)$ that have very low probability of existence.
\end{enumerate}
}

\subsection{Fusion of Bernoullis and PPPs}
\label{sec:BernoulliPPPfusion}
In this subsection, we provide expressions for the fusion of Bernoullis, fusion of PPPs, as well as the fusion of both Bernoullis and PPPs. Lastly, expressions for fusion of Gaussian densities are provided.

\subsubsection{Fusion of \acp{ppp}} \label{sec:PPPfusion}

Let $\mathbb{I}$ be an index set for PPP densities $f_{i}(\setX)$ with intensities $\lambda_i(x)=\mu_i f_{i}(x)$. Fusion of densities $f_{i}(\setX)$, $i\in\mathbb{I}$ with fusion weights $w_i$, $\sum_{i\in\mathbb{I}}w_i = 1$, yields a \ac{ppp} density with intensity \cite{clark2010robust}
\begin{subequations}
\begin{align}
    \bar{\lambda}(x) = & \prod_{i\in\mathbb{I}} \left(\lambda_{i}(x)\right)^{w_i} = \bar{\mu}\bar{f}(x), \\
    \bar{\mu} = & C \prod_{i\in\mathbb{I}}\mu_i^{w_i}   , \\
        \bar{f}(x) = & \frac{ \prod_{i\in\mathbb{I}} \left( f_{i}(x) \right)^{w_i} }{C} ,\\
        C = & \int \prod_{i\in\mathbb{I}} \left( f_{i}(x) \right)^{w_i}  {\rm d} x .
\end{align}%
\end{subequations}%

\subsubsection{Fusion of Bernoulli RFSs}
{
Consider Bernoulli densities $f_i(\setX)$ with probability of existence $r_i$ and state \ac{pdf} $f_{i}(x)$. Fusion of densities $f_{i}(\setX)$, $i\in\mathbb{I}$, $|\mathbb{I}|\geq 1$, with fusion weights $w_{i}$, $\sum_{i\in\mathbb{I}}w_{i}=1$, yields a Bernoulli density with parameters
\begin{subequations}
\begin{align}
        \bar{r} = & \frac{ C \prod_{i\in\mathbb{I}}r_i^{w_i} }{ \prod_{i\in\mathbb{I}}(1-r_i)^{w_i}  + C \prod_{i\in\mathbb{I}}r_i^{w_i} } , \\
        \bar{f}(x) = & \frac{ \prod_{i\in\mathbb{I}} \left( f_{i}(x) \right)^{w_i} }{C} ,\\
        C = & \int \prod_{i\in\mathbb{I}} \left( f_{i}(x) \right)^{w_i}  {\rm d} x .
\end{align}%
\end{subequations}%
}

\subsubsection{Fusion of Bernoulli RFSs and PPPs}
{

Let $\mathbb{I}$, $|\mathbb{I}|\geq 1$, be an index set for PPP densities $f_{i}(\setX)$ with intensities $\lambda_i(x)=\mu_i f_{i}(x)$, and let $\mathbb{J}$, $|\mathbb{J}|\geq 1$, be an index set for Bernoulli densities $f_{j}(\setX)$ with probabilities of existence $r_{j}$ and state densities $f_{j}(x)$. Fusion of PPPs and Bernoullis with fusion weights $w_{i}$ and $w_{j}$, $\sum_{i\in\mathbb{I}}w_{i}+\sum_{j\in\mathbb{J}}w_{j}=1$, yields (without approximation) a Bernoulli density with parameters
\begin{subequations}
\begin{align}
        \bar{r} = & \frac{ C \prod_{j\in\mathbb{J}}r_{j}^{w_{j}} \prod_{i\in\mathbb{I}}\mu_{i}^{w_{i}}  }{ \prod_{j\in\mathbb{J}}(1-r_{j})^{w_{j}} + C \prod_{j\in\mathbb{J}}r_{j}^{w_{j}} \prod_{i\in\mathbb{I}}\mu_{i}^{w_{i}} } , \\
        \bar{f}(x) = & \frac{ \prod_{i\in\mathbb{I}} \left( f_{i}(x) \right)^{w_{i}} \prod_{j\in\mathbb{J}} \left( f_{j}(x) \right)^{w_{j}} }{C} , \\
        C = & \int \prod_{i\in\mathbb{I}} \left( f_{i}(x) \right)^{w_{i}} \prod_{j\in\mathbb{J}} \left( f_{j}(x) \right)^{w_{j}} {\rm d} x .
\end{align}
\label{eq:BernoulliPPPfusion}
\end{subequations}

The fusion in \eqref{eq:BernoulliPPPfusion} holds for any \ac{ppp} intensities $\lambda_{i}(x)$, however, given that a Bernoulli \ac{rfs} represents zero or one object, the fusion results will be more accurate for $\mu_i<1$. }

\subsubsection{Fusion of Gaussian densities}
{
When the Bernoulli state densities, and/or the \ac{ppp} intensities are Gaussian, the fused densities $\bar{f}(\cdot)$ and the normalizing constants $C$ can be computed exactly, see, e.g., \cite[Eqn.~36]{battistelli2013consensus} and \cite{bromiley2003products}. Let $f_{i}(x) = \mathcal{N}\left(x ; m_{i}, P_{i} \right)$ for $i\in\mathbb{I}$. Fusion of the densities, with weights $w_{i}$, $\sum_{i\in\mathbb{I}}w_{i}=1$, yields a Gaussian density,
\begin{subequations}
\begin{align}
    \bar{f}(x) = & \frac{ \prod_{i\in\mathbb{I}} \left( f_{i}(x) \right)^{w_i} }{C} = \mathcal{N}\left(x ; \bar{m}, \bar{P} \right),
\end{align}
where
\begin{align}
    \bar{m} = & \bar{P} \left( \sum_{i\in\mathbb{I}} w_{i}\left( P_{i} \right)^{-1}m_{i}\right), \\
    \bar{P} = & \left( \sum_{i\in\mathbb{I}} w_{i}\left( P_{i} \right)^{-1} \right)^{-1}, \\
    C = & \int \prod_{i\in\mathbb{I}} \left( f_{i}(x) \right)^{w_i}  {\rm d} x \\
    = & \frac{ \det(2\pi\bar{P})^{1/2} }{ \prod_{i\in\mathbb{I}} \det(2\pi P_{i})^{w_{i}/2} } \exp\Bigg[ \Bigg. \\
    & \Bigg. \frac{1}{2} \left( \bar{m}^{T}\bar{P}^{-1}\bar{m} - \sum_{i\in\mathbb{I}} w_{i}m_{i}^{T}P_{i}^{-1} m_{i} \right) \Bigg]. \nonumber
\end{align}
\end{subequations}
}

\subsection{Complexity Reduction} \label{sec:bestFusionMap}
{
From \eqref{eq:KLAupperBound}, we see that we need to fuse components of the \acp{pmb} over all permutations $\pi_s\in\Pi_{(1:K)}$. The \ac{kld} is lowest when the \ac{rfs} densities are identical. In our application, this means when the single-target posterior densities represent the same target. 
To reduce computational complexity towards finding the fused posterior density minimizing the \ac{kla}, we propose to 
use the optimal permutation, denoted the \textit{best possible fusion map}.
}
It is defined and found as follows. 
Let there be two {\acp{rfs} densities} with index sets $\mathbb{I}_1$ and $\mathbb{I}_s$ respectively. We define the best fusion map $\pi_s^*\in\Pi_{(1:K)}$ as the solution of the optimal assignment problem \cite{bar1990multitarget}{
\begin{align}
\underset{\mathbf{a}}{\mathrm{minimize}}&~~\sum_{n=1}^{|\mathbb{I}_1|}\sum_{m=1}^{|\mathbb{I}_s|} a_{n,m}C_{n,m}\label{eq:assingmentProblem}\\
\mathrm{subject\; to}  &  ~~\sum_{m=1}^{|\mathbb{I}_s|} a_{n,m}=1, \;\forall n\nonumber,\\
&  ~~\sum_{n=1}^{|\mathbb{I}_1|} a_{n,m}=1, \;\forall m\nonumber,\\
& ~~a_{n,m}\in\{0,1\},\nonumber 
\end{align}}%
where $C_{n,m}$ denotes the cost for assigning (mapping) component $n$ in $f_1(\setX)$ with component $m$ in $f_s(\setX)$. To solve \eqref{eq:assingmentProblem}, we can use, e.g., the Munkres algorithm \cite{munkres1957algorithms,bourgeois1971extension}. For $\numSens>2$ the best fusion map is found by sequentially solving \eqref{eq:assingmentProblem} for $s=2,\ldots,\numSens$.

We define the cost metric  in terms of the \ac{kld} between the \acp{pdf} of the components in $f_1(\setX)$ and $f_s(\setX)$. For a component $n\in\mathbb{I}_1$ with \ac{pdf} $f_{1,n}(\statey)$, and similarly for $m\in\mathbb{I}_s$, the cost %
is defined
\begin{align}
C_{n,m} = \frac{1}{2} \left[ D(f_{1,n}\Vert f_{s,m}) + D(f_{s,m}\Vert f_{1,n}) \right],
\label{eq:costMatrixEntry}
\end{align}
which admits a closed-form expression for  Gaussian \acp{pdf}. 
Using the procedure above, we find the best fusion map and use it to solve \eqref{eq:KLAupperBound}. 

%% file: text/results.tex

We present first the target extent model used for the simulations. This is followed by the simulation setup, the used performance metrics, and a discussion of the obtained results using the independent \ac{etpmb} filter with and without posterior fusion. %
\begin{figure} 
	\centering
	\footnotesize
	\includegraphics{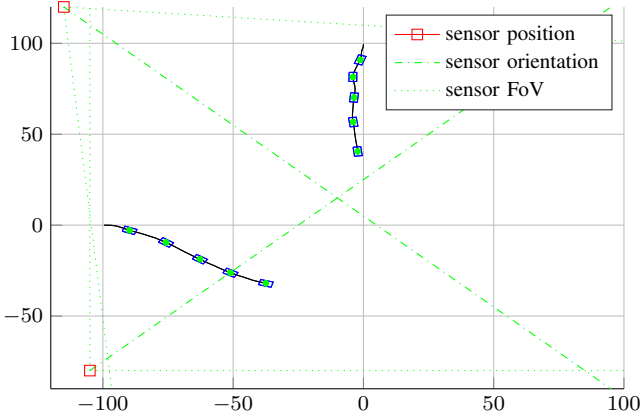}
	\caption{Simulation scenario with two \acp{et} observed by two sensors with overlapping \acp{fov}. The \acp{et} state is plot every $20$ time steps.
		\label{fig:scenario}}
\end{figure}

\subsection{Single ET Model}\label{sec:ETmodel}
The specific target extent model is based on  \cite{wahlstrom2015extended}, here extended to include the sensor state (including position and orientation) in the measurement model. In the following, we describe the \ac{et} state and measurement model, and how this leads to \ac{ekf} prediction and update equations which are utilized in the \ac{etpmb} filter. 

\subsubsection{\ac{et} State and Motion Model}
The augmented state $\statex_k=[\gamma_k,\statey_k^\mathsf{T}]^\mathsf{T}$ of a single \ac{et} at time $k$ comprises an unknown Poisson rate $\gamma_k$ of number of measurements generated by the \ac{et} and the \ac{et} spatial state $\statey_k$. The state has prior\footnote{Note that in the augmented state \ac{pdf} \eqref{eq:singleETstatePdf} the measurement rate $\gamma_k$ and the \ac{et} state (including its extent) $\statey_k$ are considered independent, which is a common assumption in \ac{ett} (see e.g.~\cite{granstrom2012estimation}). %
Due to the sensor-to-target geometry (e.g., for a Lidar its angular resolution and the sensor-target distance), the estimated measurement rate of the \ac{et} is implicitly sensor dependent, i.e., a different sensor configuration yields a different estimate for $\gamma_k$. The explicit modeling of the dependence of $\gamma_k$ on the sensor state and $\statey_k$ is not considered here. 
} \ac{pdf}
\begin{align}
f(\statex_k)=\mathcal{G}(\gamma_k;\alpha_{k},\beta_{k}) \mathcal{N}(\statey_k;\hat{\statey}_{k},\boldsymbol{P}_{k}),
\label{eq:singleETstatePdf}
\end{align}
where the gamma distribution with parameters $\alpha_{k}$ and $\beta_{k}$ is a conjugate prior for the  rate $\gamma_k$, and the 
  Gaussian distribution with mean $\hat{\statey}_{k}$ and covariance matrix $\boldsymbol{P}_{k}$ describes the a priori knowledge regarding the \ac{et} spatial state $\statey_k$. %
The \ac{et} spatial state (which includes the \ac{et} center, orientation, and extent) and motion model are based on \cite{wahlstrom2015extended} and detailed in Appendix \ref{sec:AppendixETStateAndMotion}. 
{
Since the measurement rate is independent of the sensor state, decentralized fusion of Sec.~\ref{sec:decentralizedPMBfiltering} is only applied to the spatial state $\statey_k$, while each sensor maintains a local density of the rate $\gamma_k$ (i.e., average number of measurements obtained from target with specific sensor) of each target. }

\subsubsection{\ac{et} Measurement Model} \label{sec:ETMeasurementModel}
In this section, the time index $k$ will be omitted for the sake of brevity.
A sensor $s$ located at $\boldsymbol{p}_s\in\mathbb{R}^2$ with orientation $\alpha_s$ observes the \ac{et} contour in its local coordinate frame. We distinguish between three coordinate frames: a quantity in the \textit{sensor coordinate frame} is indicated by superscript $S$, in the \textit{\ac{et} coordinate frame} by superscript $L$, and in the \textit{global coordinate frame} by superscript $G$. A measurement $\boldsymbol{z}^S\in\mathbb{R}^2$ is thus
\begin{align}
\boldsymbol{z}^S = h(\statey) + \boldsymbol{w}, \label{eq:zS}
\end{align}
where $\boldsymbol{w}\sim\mathcal{N}(0,R)$ with measurement noise covariance $R$ and 
\begin{align}
h(\statey)= \statey^S_c + \boldsymbol{e}(\theta^S)^\mathsf{T} f(\theta^L),
\label{eq:hy}
\end{align}
where $\theta^S=\angle(\boldsymbol{z}^S-\statey^S_c)$, $\boldsymbol{e}(\theta^S)$ is a unit vector in direction $\theta^S$, and $\theta^L=\theta^S-\psi^S$ in which $\statey^S_c$ and $\psi^S$ are the target location and orientation in the sensor frame of reference.\footnote{Note that here the unknown angle $\theta^S$ is replaced by a point estimate, which is a simple but inaccurate approach and can be seen as a greedy association model \cite{granstromETT2017}. It is also the approach taken by \cite{wahlstrom2015extended}.} Here, $ f(\theta^L)\ge 0$ is the extent of the target along local angle $\theta^L$. We now express the observation in the global coordinate frame. It can  be shown that $\boldsymbol{z}^S \sim \mathcal{N}(\tilde{h}(\statey),\tilde{R})$ with
\begin{align}
\tilde{h}(\statey) & = R(\alpha_s)(\statey_c-\boldsymbol{p}_s) 
+ \boldsymbol{e}^\mathsf{T} (\theta^S) H^{\statef}(\theta^L)\statey^{\boldsymbol{f}}  \label{eq:zS2},\\
\tilde{R} & =R+ k^f \boldsymbol{e}(\theta^S)\boldsymbol{e}^\mathsf{T}(\theta^S),
\label{eq:zS3}
\end{align} 
where $R(\alpha_s)$ is a rotation matrix, $ H^{\statef}(\theta^L)$, and $k^f$ are defined in Appendix \ref{sec:AppendixObsModel}.

\subsubsection{ET Extended Kalman Filter Equations}
Prediction and update equations of an \ac{ekf} filter using the ET motion and measurement models are given in Appendix~\ref{sec:appendixEKFpredupdate}.  The prediction and update equations for the measurement rate and the predicted likelihood are also stated.
All these are utilized in the prediction and update step of the \ac{etpmb} filter (c.f.~Section~\ref{sec:centralizedPMBFilter}).

\subsection{Setup}

If not stated otherwise, there are two \acp{et} present in the scene and their visibility and number of measurements produced per scan depends on the sensor \ac{fov} and its configuration. We use a rectangular target of length $5~\mathrm{m}$ and width $3~\mathrm{m}$ to model \acp{et} representing vehicles. 
Furthermore, we use Lidar type sensors with the following simplified sensor models. Sensor 1 is located at $\boldsymbol{p}_{S_1}=[-115,120]^\mathsf{T}$, with orientation $\alpha_{S_1}=-45^{\circ}$, opening angle of $80^{\circ}$, angular resolution of $0.15^{\circ}$, and maximum range of $300~\mathrm{m}$. We generate a measurement when a ray from the sensor hits an \ac{et} and add noise with covariance matrix $R_{S_1}=0.5\boldsymbol{I}_2$. Sensor 2 is located at $\boldsymbol{p}_{S_2}=[-105,-80]^\mathsf{T}$, with orientation $\alpha_{S_2}=45^{\circ}$, opening angle of $90^{\circ}$, angular resolution of $0.15^{\circ}$, maximum range of $300~\mathrm{m}$, and $R_{S_2}=0.02\boldsymbol{I}_2$. Each sensor produces clutter measurements with rate $\lambda=2$ and uniform spatial distribution $c(\statez)=\mathcal{U}[-200,200]^2$.

In the simulation, each independent \ac{etpmb} filter has only access to measurements from one sensor (denoted indep. filter). The fusion filters are independent \ac{etpmb} filters (denoted fusion filter), but perform posterior fusion according to Sec.~\ref{sec:decentralizedPMBfiltering}. We approximate the \ac{pmbm} posterior by a \ac{pmb} which consists of the \ac{mb} in the \ac{mbm} that has highest weight.
If not stated otherwise, posterior fusion is applied in every time step. We set the weights $\omega_s=1/\numSens$, resulting in a more conservative estimate than achievable. %
The fused posterior is then used as the prior for the next filter iteration. For comparison, an \ac{etpmb} filter which incorporates measurements from all sensors is used (denoted centralized filter). There, measurements from each sensor are incorporated separately by applying multiple sequential \ac{etpmb} filter update steps.

For all filter variants, spatially close measurements are clustered into measurement cells  using the DBSCAN clustering algorithm \cite{ester1996density}, where we set the maximum radius for the neighborhood to $4~\mathrm{m}$ and the minimum number of points for a core point to $4$. The simulation scenario is outlined in Fig.~\ref{fig:scenario}.
The hyper parameters of the \ac{gp} (see Appendix~\ref{sec:AppendixETStateAndMotion}) are $l^2=\pi/8$, $\sigma^2_f=2$, $\sigma^2_r=2$, and 20 support points are used to track the target extent. Note that the dimension of the \ac{et} state is 26 ($xy$-position, orientation, $xy$-velocity, angular velocity, target extent support points). 
Target motion model and its model in the filter is identical with sampling time $T=0.5~\mathrm{s}$, 
\begin{align}\bar{F}_k=&\begin{bmatrix}
1 &T \\0 &1 
\end{bmatrix}\otimes \boldsymbol{I}_3,\\
\bar{W}_k=&\begin{bmatrix}
\frac{T^3}{3} & \frac{T^2}{2}\\
\frac{T^2}{2} & T
\end{bmatrix}\otimes \mathrm{diag}([0.01,0.01,0.001]),
\end{align}
$\beta = 0.001$ in $F_{\boldsymbol{f}}$ (c.f.~\eqref{eq:ETmotionModel}), and the forgetting factor is set to $\frac{1}{\eta}=\frac{1}{1.11}$. In the filters, the birth intensity has rate $\lambda^b=\frac{1}{10}$ and for the spatial distribution we use a single Gaussian centered at location $x=[0,100]^\mathsf{T}$ with covariance matrix $P=30\boldsymbol{I}_2$. The probability of \ac{et} survival is $p_\mathrm{S}=0.999$, the probability of detection is $p_\mathrm{D}=0.99$. 
\begin{figure}
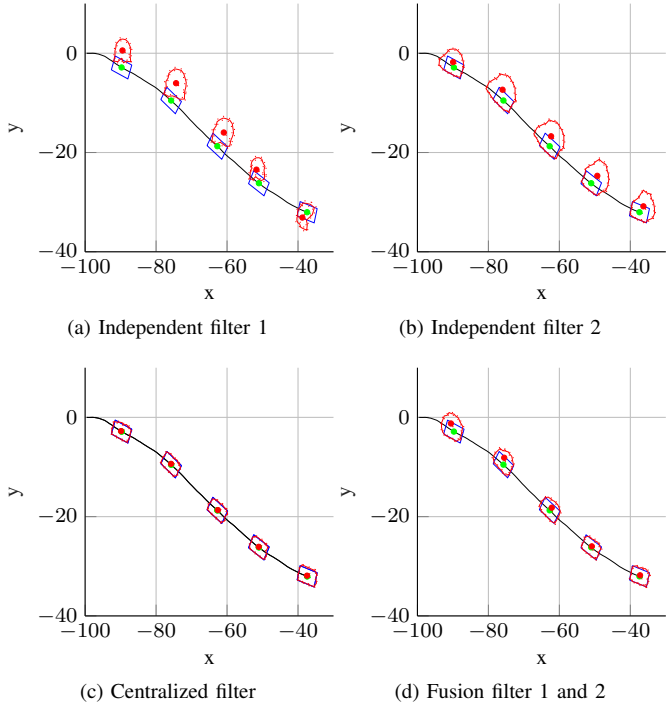
 
	\centering
	\footnotesize
	\subfloat[Independent filter 1]{%
		\label{fig:out_independentS1}%
		\includegraphics{figures/out_independentS1_v2.tikz}} %
	\subfloat[Independent filter 2]{%
		\label{fig:out_independentS2}%
		\includegraphics{figures/out_independentS2_v2.tikz}} %
	\\
	\subfloat[Centralized filter]{%
		\label{fig:out_centralized}%
		\includegraphics{figures/out_centralized_v2.tikz}} %
	\subfloat[Fusion filter 1 and 2]{%
		\label{fig:out_fusedS1}%
		\includegraphics{figures/out_fusedS1_v2.tikz}} %
	\caption{Estimated shape of one \ac{et} for different \ac{etpmb} filters. The true \ac{et} center and extent is plotted (green dot, blue solid line), as well as the estimated ones (red dot, red solid line for mean extent, red dotted line for one standard deviation). }
	\label{fig:out_plots}
\end{figure}
\subsection{Performance Metrics}

{Multiple target tracking performance is measured by three errors: estimation error for localized targets, number of missed targets, and number of false targets, see, e.g.,  \cite[Sec. 13.6]{BlackmanP:1999}. The \ac{gospa} metric \cite{rahmathullah2016metric,rahmathullah2017generalized} measures all three errors, hence we use it for performance evaluation.\footnote{In tracking literature, the OSPA metric \cite{SchuhmacherVV:2008} is often used. However, recent work \cite{GarciaFernandezS:2019} has shown that  the OSPA metric is susceptible to ``spooky action at a distance'', which is undesirable. Hence, we do not use the OSPA metric in this paper.}}
Performance of the estimated number of targets as well as their center location is thus assessed as follows. Let sets $\hat{\setX}=\{\statex_1,\ldots,\statex_n\}$ and $\setY=\{\statey_1,\ldots,\statey_m\}$ be finite subsets of $\mathbb{R}^N$, where without loss of generality $n \leq m$. Then \cite{rahmathullah2016metric,rahmathullah2017generalized}
\begin{align}
d_\mathrm{GOSPA}^{(c,\alpha,p)} =& \left( \min_{I_n\in\mathcal{F}_n(\{1,\ldots,n\})} 
\sum_{i=1}^n d^{(c)}(\statex_i,\statey_{I_n(i)})^p \right.\nonumber\\
& + \left.
\frac{c^p}{\alpha} (m-n)
\right)^{\frac{1}{p}},
\end{align}
where we set the power parameter $p=2$, cut-off distance $c=20$,  $\alpha=2$,
and $d^{(c)}(\statex,\statey)=\min(\Vert\statex-\statey\Vert_2,c)$, i.e., the minimum of the Euclidean distance and value $c$.
To obtain $\hat{\setX}$, we estimate the detected \acp{et} from the (fused) posterior through comparison of the probability of existence of each Bernoulli component against the threshold $r_\mathrm{th}=0.5$. The set $\setY$ contains the true \acp{et}.
Note that we only use the mean position of the target center (c.f. Sec.~\ref{sec:bayesianFilterFormulation}).

Performance of the target extent estimation is assessed with the \ac{iou} of the true target shape (in the global coordinate frame) and the estimated shape. Let $A_k$ be the true \ac{et} area in $xy$-dimension at time step $k$, and $\hat{A}_k$ its estimate. Then, the \ac{iou} is defined as, see, e.g., \cite{wahlstrom2015extended},
\begin{align}
\mathrm{IOU}(A_k,\hat{A}_k) = \frac{\mathrm{area}(A_k\cap \hat{A}_k)}{\mathrm{area}(A_k\cup\hat{A}_k)}.
\end{align}
Note that the \ac{iou} is, by definition, always between zero for non-overlapping target shapes and one when they fully overlap. Thus a well performing \ac{ett} filter will yield a high \ac{iou} for every \ac{et}. For both \ac{gospa} and \ac{iou} average performance results were obtained by averaging over $50$ Monte-Carlo runs.

\begin{figure*}
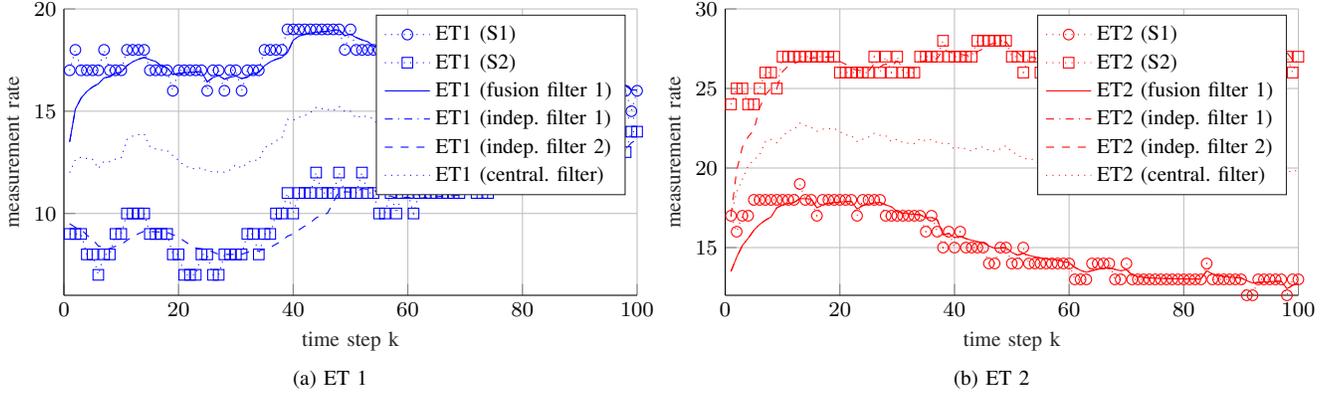
 
	\centering
	\footnotesize
	\subfloat[\ac{et} 1]{%
		\label{fig:rateET1}%
		\includegraphics{figures/out_rate_ET1_v2.tikz}} %
	%
	\subfloat[\ac{et} 2]{%
		\label{fig:rateET2}%
		\includegraphics{figures/out_rate_ET2_v2.tikz}} %
	\caption{The true and estimated measurement rate for each \ac{et} and sensor/filter is plotted over time.
		\label{fig:rate}}
\end{figure*}

\subsection{Discussion of Results}
Here, we first discuss the \ac{ett} filter performance with decentralized posterior fusion in terms of \ac{gospa} and \ac{iou} performance metric. This is followed by performing posterior fusion at a lower rate. After that, we investigate the case when more than two \ac{ett} filters are used for posterior fusion.

\subsubsection{Filter Performance with Posterior Fusion}

In Fig.~\ref{fig:out_plots}, the true and estimated target state is plotted for \ac{et} 2 using the different filter variants. We observe that independent filter 1 (Fig.~\ref{fig:out_independentS1}) estimates the target state with a clear position error (in the positive $y$-direction). 
This filter uses measurements from sensor 1, where most of the measurements provide information only from the \ac{et}'s top edge due to the horizontal movement of the target and the sensor pose.
In the measurement model \eqref{eq:zS}, occlusions caused by the \ac{et} itself are not modeled. 
Due to the simple data association that is used, the received measurements can then be explained by an \ac{et} whose target contour is in proximity of the measurements and the target center is placed north of it.
 In contrast, independent filter 2 and the centralized filter 
estimate the target center closer to the true position (Fig.~\ref{fig:out_independentS2} and Fig.~\ref{fig:out_centralized}).
The former filter overestimates the target size in the direction where no measurement is provided, whereas the latter filter utilizes measurements from both sensors and can therefore estimate the target size accurately.
The fusion filter utilizes information from both sensors through posterior fusion. 

In Fig.~\ref{fig:rate}, the true and estimated measurement rate $\gamma$ is plotted over time. We can observe that the true measurement rate of the \acp{et} is varying over time. Although only a simple process model for the measurement rate is used in the filters, they are able to correctly track the rate. This is true for all filter variants except the centralized filter. This filter performs two sequential filter update steps using measurement from different sensors. Since the number of measurements for an \ac{et} are different for each sensor, it follows that after centralized fusion the filter estimates the measurement rate of the \ac{et} as the average of the two.

\begin{figure} 
	\centering
	\footnotesize
	\includegraphics{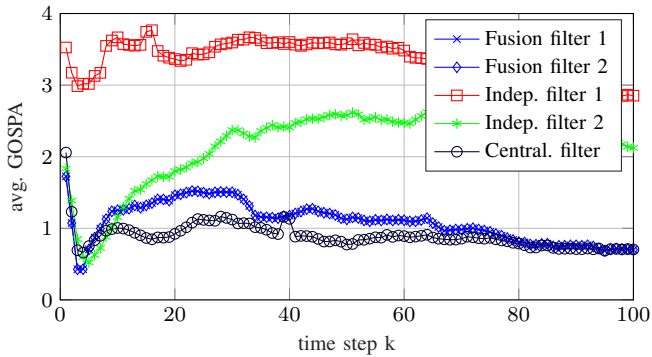} 
	\caption{The average \ac{gospa} value is plotted over time. Posterior fusion is performed in every time step.
		\label{fig:gospa}}
\end{figure}

In Fig.~\ref{fig:gospa}, the average \ac{gospa} is plotted over time for the simulation scenario illustrated in Fig.~\ref{fig:scenario}.
The centralized filter has best performance followed by the fusion filter. Independent filter 2 has superior performance compared to the independent filter 1. 
Note that independent filter 1 is provided with measurements from sensor 1, which has a higher measurement noise compared to sensor 2. %

In Table~\ref{tab:iou}, the average \ac{iou} is stated for the different \acp{et} and filter variants. 
We observe that the \ac{iou} for \ac{et} 2 is low with independent filter 1 due to the misplaced target center, and with independent filter 2 due to the overestimation of the target size.
The fusion and the centralized filter show comparable performance. 

\begin{table}[h!]
	\caption{ Average \ac{iou}  }
	\centering
	\begin{tabular}{|c|c|c|}
		\hline 
		Filter & \ac{et} 1 & \ac{et} 2\tabularnewline
		\hline 
		\hline 
		Independent 1 &  0.65 & 0.25 \tabularnewline
		\hline 
		Independent 2 & 0.71 & 0.43 \tabularnewline
		\hline 
		Fusion & 0.72 & 0.75 \tabularnewline
		\hline 
		Centralized & 0.70 & 0.85 \tabularnewline
		\hline 
	\end{tabular}\label{tab:iou}
\end{table}

\subsubsection{Low Rate Posterior Fusion}
In a real system, it may not be feasible to perform posterior fusion after every filter update step. This can occur when the computers on which the filters run are geographically separated and need to communicate over the wireless channel. Therefore, it is worthwhile to investigate the filter performance when posterior fusion is performed only every $N$ time steps. 

In Table~\ref{tab:gospa}, the average \ac{gospa} is stated for different values of $N$. We see that with increasing $N$ the performance of the fusion filters deteriorates, since the information transfer (through fusion) between the filters over time is too low. Fusion filter 1 has worse performance compared to fusion filter 2 for $N\geq 15$, since it is equipped with the low quality sensor 1.

\begin{table}[h!]
	\caption{ Average \ac{gospa}  }
	\centering
	\begin{tabular}{|c|c|c|c|c|}
		\hline 
		Filter & $N=1$ & $N=15$ & $N=30$ & $N=50$\tabularnewline
		\hline 
		\hline
		Fusion 1 & 1.12 & 1.65 & 2.20 & 2.76 \tabularnewline
		\hline
		Fusion 2 & 1.12 & 1.80 & 2.03 & 2.10 \tabularnewline
		\hline 
		Independent 1 & 3.20 & 3.20 & 3.20 & 3.20 \tabularnewline
		\hline 
		Independent 2 & 2.17 & 2.17 & 2.17 & 2.17 \tabularnewline
		\hline 
		Centralized & 0.89 & 0.89 & 0.89 & 0.89 \tabularnewline
		\hline 
	\end{tabular}\label{tab:gospa}
\end{table}
\subsubsection{Posterior Fusion with more filters}
In Sec.~\ref{sec:decentralizedPMBfiltering}, we proposed a procedure to fuse \ac{pmb} posteriors when there are more than two independent \ac{etpmb} filters. We implement this sequentially, where first posteriors from two filters are fused. The outcome is then used to fuse with a not yet fused posterior from one of the remaining filters. This process is repeated until all filter posteriors have been incorporated.
We placed four sensors at $\boldsymbol{p}_{S_1}=[-150,-80]^\mathsf{T}$, $\boldsymbol{p}_{S_2}=[-150,-50]^\mathsf{T}$, $\boldsymbol{p}_{S_3}=[-150,-20]^\mathsf{T}$, and $\boldsymbol{p}_{S_4}=[-150,10]^\mathsf{T}$ all with overlapping sensor \acp{fov} towards the \acp{et}. The remaining sensor parameters are the same as for sensor 1 used in the previous simulations. 
In Table~\ref{tab:MultiFusion}, the average \ac{gospa}, as well as the average \ac{iou} per \ac{et} are stated for a single independent \ac{etpmb} filter (no fusion), two filters with posterior fusion performed after every filter update, and four filters with posterior fusion. With posterior fusion, performance increases, visible by a decrease of the \ac{gospa} value. Also the average \ac{iou} increases for all \acp{et} with posterior fusion. Furthermore, fusion of two filter posteriors and four filter posteriors show similar performance in this scenario.

\begin{table}[h!] 
	\caption{ Multiple Posterior Fusion  }
	\centering
	\begin{tabular}{|c|c|c|c|}
		\hline 
		No. Posteriors Fused & \ac{gospa} & IOU (\ac{et} 1) & IOU (\ac{et} 2) \tabularnewline
		\hline 
		\hline
		(no fusion) & 1.22 & 0.61 & 0.62 \tabularnewline
		\hline
		2 & 0.90 & 0.66 & 0.67 \tabularnewline
		\hline 
		4 & 0.86 & 0.68 & 0.66  \tabularnewline
		\hline 
	\end{tabular}\label{tab:MultiFusion}
\end{table}

%% file: text/conclusions.tex

We proposed a low-complexity decentralized \ac{ett} filter that is capable of estimating the presence, state, and shape of \acp{et} accurately. It operates by combining the multibobject densities of \ac{pmb} form computed by independent \ac{ett} filters. 
Fusion is performed in the minimum \ac{kla} sense yielding a fused posterior which is conservative but never overconfident about the estimated states. A low-complexity implementation is highlighted,
which permits the use of an optimal fusion mapping between pairs of sensors. The fusion map was identified as the solution of a linear optimal assignment problem based on a cost matrix comprised of the symmetric \ac{kld} between target state estimates.

In the simulation results, we observed how the independent \ac{etpmb} filters together with the \ac{gp} target extent model are capable of estimating the state and shape of the present \acp{et}. Furthermore, we observed that fusion of the filters posterior permits a holistic view of the surveillance region spanned by all sensors combined. This resulted in a reduced state estimation error quantified by the \ac{gospa} distance metric, and for the \ac{et} shape estimation by an increased area overlap quantified by the \ac{iou} value. 

%% file: text/appendix.tex

\section{ET-PMB parameters} \label{sec:Appendixetpmb}
Following \cite[Sec.~IV]{GranstromFS:2016_PMBMETT}, we find that 
\begin{align}
w_\assoc =& \frac{\prod_{C\in\assoc}\mathcal{L}_C}{ \sum_{\assoc\in\assocspace} \prod_{C\in \assoc} \mathcal{L}_C },\\
\mathcal{L}_C = &
\begin{cases}
\kappa^{\setC_C} + \conv{D_+^u}{\ell_{\setC_C}}, 	& \mathrm{if}\; C \cap \mathbb{I}=\emptyset,|\setC_C|=1,  \\
\conv{D_+^u}{\ell_{\setC_C}}, 				& \mathrm{if}\; C \cap \mathbb{I}=\emptyset,|\setC_C|>1,  \\
1-r_+^{i_C}+r_+^{i_C}\conv{f_+^{i_C}}{q_\mathrm{D}}, & \mathrm{if}\; C \cap \mathbb{I}\neq\emptyset,\setC_C=\emptyset,  \\
r_+^{i_C}\conv{f_+^{i_C}}{\ell_{\setC_C}}, 	& \mathrm{if}\; C \cap \mathbb{I}\neq\emptyset,\setC_C\neq\emptyset.
\end{cases}
\end{align}
The density $f_C(\setX^C)$ in \eqref{eq:PMBFilterfd} is a Bernoulli density with parameters
\begin{align}
r_C = &
\begin{cases}
\frac{\conv{D_+^u}{\ell_{\setC_C}}}{\kappa^{\setC_C} + \conv{D_+^u}{\ell_{\setC_C}}}, 	& \mathrm{if}\; C \cap \mathbb{I}=\emptyset,|\setC_C|=1,  \\
1,																						& \mathrm{if}\; C \cap \mathbb{I}=\emptyset,|\setC_C|>1,  \\
\frac{r_+^{i_C}\conv{f_+^{i_c}}{q_\mathrm{D}}}{1-r_+^{i_C}+r_+^{i_C}\conv{f_+^{i_C}}{q_\mathrm{D}}},		& \mathrm{if}\; C \cap \mathbb{I}\neq\emptyset,\setC_C=\emptyset,  \\
1, 																						& \mathrm{if}\; C \cap \mathbb{I}\neq\emptyset,\setC_C\neq\emptyset,
\end{cases}
\end{align}
\begin{align}
f_C(\statex) =&
\begin{cases}
\frac{\ell_{\setC_C}D_+^u(\statex)}{\conv{D_+^u}{\ell_{\setC_C}}},			& \mathrm{if}\; C \cap \mathbb{I}=\emptyset,  \\
\frac{q_D(\statex)f_+^{i_C}(\statex)}{\conv{f^{i_C}}{q_\mathrm{D}}},					& \mathrm{if}\; C \cap \mathbb{I}\neq\emptyset,\setC_C=\emptyset,  \\
\frac{\ell_{\setC_C}(\statex)f_+^{i_C}(\statex)}{\conv{f_+^{i_C}}{\ell_{\setC_C}}}, & \mathrm{if}\; C \cap \mathbb{I}\neq\emptyset,\setC_C\neq\emptyset.
\end{cases}
\end{align}

\section{ET Spatial State and Motion Model} \label{sec:AppendixETStateAndMotion}
\subsection{Spatial State}

The \ac{et} spatial state including target extent is given by
\begin{equation}
\statey=[(\bar{\statey})^\mathsf{T}, (\statey^{\boldsymbol{f}})^\mathsf{T}]^\mathsf{T},
\end{equation}
where
\begin{equation}
\bar{\statey} = [(\statey_c)^\mathsf{T}, \psi, {\statey^*}^\mathsf{T}]^\mathsf{T}
\end{equation}
comprising 
 the \ac{et} center $\statey_c$, the \ac{et} orientation $\psi$, and any additional quantities (e.g, velocity) in $\statey^*$. The variable $\statey^{\boldsymbol{f}}$ models the target extent, following 
 \cite{wahlstrom2015extended}: 
 Let $u$ denote the local angle \ac{wrt} the \ac{et} orientation and $\stateyScal^{f}_i$ denotes the unknown target extent along input (angle) $u_i$, for a fixed and finite set of $N$ angles. Then %
\begin{equation}
\statey^{\boldsymbol{f}}=[\stateyScal^{f_1},\ldots,\stateyScal^{f_N}]^\mathsf{T}.
\end{equation}
This vector is modeled as a zero mean \ac{gp} \cite{williams2006gaussian,bishop2006pattern}
\begin{equation}
\statey^{\boldsymbol{f}}\sim\mathcal{GP}(\boldsymbol{0},K(\boldsymbol{u},\boldsymbol{u}))
\label{eq:zeroMeanGPModel}
\end{equation}
with  covariance matrix $[K(\boldsymbol{u},\boldsymbol{u})]_{i,j}=k(u_i,u_j)$, 
in which $k(\cdot,\cdot)$ is a periodic kernel function. The input of the \ac{gp} is $\boldsymbol{u}=[u_1,\ldots,u_N]^\mathsf{T}$, and the output is $\statey^{\boldsymbol{f}}$. {We utilize the periodic kernel function proposed in \cite{wahlstrom2015extended}
\begin{align}
k(u,u')=\sigma_{f}^2 \exp\left( -\frac{2\sin^2\left( \frac{\vert u-u'\vert}{2} \right)}{l^2}  \right) + \sigma_r^2,
\end{align}
where $\sigma_f$, $l$ and $\sigma_r$ are the (known) model hyper-parameters. This function is $2\pi$ periodic, i.e., $k(u+2\pi,u')=k(u,u')$, and, thanks to $\sigma_r$, star convex object shapes of different sizes can be described. See \cite{wahlstrom2015extended,hirscher2016multiple} for further details and different choices for the kernel function to describe the extent of an \ac{et} with the help of a \ac{gp}.}

\subsection{Motion Model}\label{sec:ETmotionmodel}

Extending Sec.~\ref{sec:ETstatetransitionmodel}, the \ac{et} follows the linear dynamic model 
\begin{align}
\bar{\statey}_{k+1} = \bar{F}_k\bar{\statey}_k+\bar{\boldsymbol{w}}_k,
\end{align}
where $\bar{F}_k$ denotes the state transition matrix, and $\bar{\boldsymbol{w}}_k\sim\mathcal{N}(0,\bar{W}_k)$ with process noise covariance $\bar{W}_k$.
The motion model of the \ac{et} contour is \cite{wahlstrom2015extended}
\begin{align}
\statey^{\boldsymbol{f}}_{k+1} = F^{\boldsymbol{f}}_k\statey^{\boldsymbol{f}}_k + \boldsymbol{w}^{\boldsymbol{f}}_k,
\label{eq:ETmotionModel}
\end{align}
where $F^{\boldsymbol{f}}=e^{-\beta T}\boldsymbol{I}_N$ with $\boldsymbol{I}_N$ denoting the identity matrix of dimension $N$, and $\boldsymbol{w}^{\boldsymbol{f}}_k\sim\mathcal{N}(0,W^{\boldsymbol{f}})$ with $W^{\boldsymbol{f}}=(1-e^{-2\beta T})K(\boldsymbol{u}^{\boldsymbol{f}},\boldsymbol{u}^{\boldsymbol{f}})$. Here, $\beta\geq 0$ denotes the {forgetting} factor allowing to accommodate targets with extents that change slowly, and $T$ is the sampling time.

According to \eqref{eq:singleETstatePdf}, the measurement rate of the \ac{et} is assumed independent of the \acp{et}' spatial state. %
To allow the measurement rate to change over time an exponential forgetting factor ${1}/{\eta}$ is used in the motion model, where the predicted rate is given by the motion of the gamma distribution parameters with \cite{granstrom2012estimation, GranstromFS:2016_PMBMETT}
\begin{align} 
\alpha_{k|k-1} &= \alpha_{k-1}/\eta,\\
\beta_{k|k-1} &= \beta_{k-1}/\eta. \label{eq:predictedRate}
\end{align}

\section{Proof of (49)--(50)}\label{sec:AppendixObsModel}

Due to the GP model, we follow \cite{wahlstrom2015extended} and express $f(\theta^L)$ as 
\begin{align}
f(\theta^L)=& H^{\statef}(\theta^L) \statey^{\boldsymbol{f}}+e^f,\\
H^{\statef}(\theta^L) =& K(\theta^L,\boldsymbol{u}^{\boldsymbol{f}})[K(\boldsymbol{u}^{\boldsymbol{f}},\boldsymbol{u}^{\boldsymbol{f}})]^{-1},
\label{eq:GPmean}
\end{align}
in which $e^f \sim \mathcal{N}(0,k^f)$ with 
\begin{align}
k^f = k(\theta^L,\theta^L)-H^{\statef}(\theta^L) K(\theta^L,\boldsymbol{u}^{\boldsymbol{f}})^\mathsf{T}. 
\label{eq:GPvar}
\end{align}

Separating signal and noise contribution, and expressing the local states in global frame of reference through 
\begin{align}
\statey^S_c & = R(\alpha_s)(\statey_c-\boldsymbol{p}_s)\\
\psi^S & = \psi- \alpha_s, 
\end{align}
where $R(\alpha)$ denotes the rotation matrix
\begin{align}
R(\alpha) = \begin{bmatrix}
\cos(\alpha) & -\sin(\alpha)\\
\sin(\alpha) & \cos(\alpha)
\end{bmatrix},
\end{align}
we find that with the \ac{gp} contour model measurement $\boldsymbol{z}^S$ in \eqref{eq:zS} has mean and covariance given by \eqref{eq:zS2}--\eqref{eq:zS3}.

\section{ET prediction and update steps} \label{sec:appendixEKFpredupdate}
Here, we first describe the \ac{ekf} prediction and update steps for the \ac{et}'s spatial state. This is followed by the update step of the \ac{et} measurement rate, and lastly the predicted likelihood utilized in the update step of the \ac{etpmb} filter for the \ac{et} state model of Section~\ref{sec:sysModelAndProbFormulation}.
\subsection{EKF prediction and update equations}
With the linear \ac{et} motion model the standard \ac{ekf} prediction step with initial state $\statey_0\sim\mathcal{N}(\hat{\statey}_0,P_0)$ is \cite{simon2006optimal}
\begin{align}
\hat{\statey}_{k|k-1} =& F_k \hat{\statey}_{k-1},\\
P_{k|k-1} = & F_k P_{k-1} F_k^\mathsf{T} + W_k,
\end{align}
where $F_k=\mathrm{blkdiag}(\bar{F}_k, F_k^{\boldsymbol{f}})$, and $W_k = \mathrm{blkdiag}(\bar{W}_k, W_k^{\boldsymbol{f}})$.

We now extend the \ac{ekf} update steps derived in \cite{wahlstrom2015extended} to incorporate the (known) sensor state (position $\boldsymbol{p}_s$ and orientation $\alpha_s$). The standard \ac{ekf} measurement update equations for a detection $\statez_k$ are \cite{simon2006optimal}
\begin{align}
H_k &= \frac{\diff}{\diff\statey_k}{\tilde{h}}(\statey)|_{\statey=\hat{\statey}_{k|k-1}},\\
S_k&=H_k P_{k|k-1}H_k^\mathsf{T}+R_k,\\
K_k&=P_{k|k-1}H_k^\mathsf{T}S_k^{-1},\\
\hat{\statey}_{k}&=\hat{\statey}_{k|k-1}+K_k (\statez_k - \tilde{h}(\hat{\statey}_{k|k-1})),\\
P_{k}&=P_{k|k-1}-K_k H_k P_{k|k-1},
\end{align}
where $\tilde{h}(\cdot)$ was defined in \eqref{eq:zS2} and \eqref{eq:zS3}.
To linearize the measurement function, we need to compute \cite{wahlstrom2015extended}
\begin{align}
H_k = \begin{bmatrix}
{\frac{\diff \tilde{h}(\statey)}{\diff \statey_c}},
{\frac{\diff \tilde{h}(\statey)}{\diff \psi}},
{\frac{\diff \tilde{h}(\statey)}{\diff \statey^*}},
{\frac{\diff \tilde{h}(\statey)}{\diff \statey^f}}
\end{bmatrix},
\end{align}
where in our case ${\frac{\diff \tilde{h}(\statey)}{\diff \statey^*}}=0$.
We get
\begin{align}
\frac{\diff \tilde{h}(\statey)}{\diff \statey^{\statef}} =& \boldsymbol{e}(\theta^S)H^{\statef}(\theta^L),\\
\frac{\diff \tilde{h}(\statey)}{\diff \psi} =& \boldsymbol{e}(\theta^S)\frac{\diff}{\diff\psi}H^{\statef}(\theta^L)\statey^{\statef},\\
\frac{\diff}{\diff\psi}H^{\statef}(\theta^L) =&
-\left.\frac{\partial}{\partial u}H^{\statef}(u)\right\vert_{u=\theta^L},
\end{align}
where \cite{wahlstrom2015extended}
\begin{align}
\frac{\diff H^{\statef}(u)}{\diff u} = &\frac{\diff }{\diff u}
K(u,\boldsymbol{u}^{\boldsymbol{f}}) [K(\boldsymbol{u}^{\boldsymbol{f}},\boldsymbol{u}^{\boldsymbol{f}})]^{-1},\\
\frac{\diff K(u,\boldsymbol{u}^{\boldsymbol{f}})}{\diff u} = &
\frac{\diff}{\diff u}[k(u,u_1^{\boldsymbol{f}}),\ldots,k(u,u_N^{\boldsymbol{f}})],\\
\frac{\diff k(u,u_i^{\boldsymbol{f}})}{\diff u}=&
-\frac{1}{l^2}\sin(u-u_i^{\boldsymbol{f}})k(u,u_i^{\boldsymbol{f}}).
\end{align}
Further,
\begin{align}
&\frac{\diff \tilde{h}(\statey)}{\diff \statey_c} = R(\alpha_s)+\left.\frac{\diff}{\diff \boldsymbol{u}}\boldsymbol{e}(\boldsymbol{u})\right\vert_{\boldsymbol{u}=\statey_c} (H^{\statef}(\theta^L)\statey^{\statef})^\mathsf{T} 
+  \boldsymbol{e}(\statey_c) \nonumber\\
&\times\left( \left( \left(\left.\frac{\diff}{\diff u}H^{\statef}(u)\right\vert_{u=\theta^L} \right)^\mathsf{T}
\left.\frac{\diff}{\diff \boldsymbol{w}}\theta^L(\boldsymbol{w})\right\vert_{\boldsymbol{w}=\statey_c}\right)^\mathsf{T} \statey^{\statef}\right)^\mathsf{T},
\end{align}
where \cite{wahlstrom2015extended}
\begin{align}
\frac{\diff}{\diff \boldsymbol{u}}\boldsymbol{e}(\boldsymbol{u}) = &\frac{(\statez^S- \boldsymbol{u})(\statez^S- \boldsymbol{u})^\mathsf{T}}{\Vert\statez^S- \boldsymbol{u}\Vert^3} - \frac{1}{\Vert\statez^S- \boldsymbol{u}\Vert} \boldsymbol{I}_2,\\
\frac{\diff}{\diff \boldsymbol{w}}\theta^L(\boldsymbol{w}) = &
\frac{1}{\Vert \boldsymbol{z}^S - \statey_c^S \Vert^2}\nonumber\\
&\times\begin{bmatrix}
{\boldsymbol{z}^S}^Y-{\statey_c^S}^Y, -({\boldsymbol{z}^S}^X-{\statey_c^S}^X)
\end{bmatrix} R(\alpha_s).
\end{align}
Here, the superscript $\boldsymbol{b}^X$ and $\boldsymbol{b}^Y$ correspond to the first and second dimension of the vector $\boldsymbol{b}$. Furthermore, we wrote $\boldsymbol{e}(\statey_c)$ for $\boldsymbol{e}(\theta^S)$ to indicate the state dependency.

To update the \ac{et} spatial state by a set of detections $\boldsymbol{W}=\left.\{\boldsymbol{z}_{k,l}\}\right._{l=1}^{n_k}$, we augment the measurement vector 
\begin{align}
\boldsymbol{z}_k &=[\boldsymbol{z}_{k,1}^\mathsf{T},\ldots,\boldsymbol{z}_{k,n_k}^\mathsf{T}]^\mathsf{T},
\end{align}
where
\begin{align}
R_k &= \mathrm{diag}(R_{k,1},\ldots,R_{k,n_k}),\\
\tilde{h}_k(\statey_k)&=[\tilde{h}_{k,1}(\statey_k)^\mathsf{T},\ldots,\tilde{h}_{k,n_k}(\statey_k)^\mathsf{T}]^\mathsf{T}.
\end{align}
\subsection{ET measurement rate}
The predicted \ac{et} measurement rate $\gamma_{k|k-1}$ has parameters $\alpha_{k|k-1}$ and $\beta_{k|k-1}$ (c.f.~\eqref{eq:singleETstatePdf}, \eqref{eq:predictedRate}), which are updated for a set of detections $\mathbf{W}$ by \cite{GranstromFS:2016_PMBMETT}
	\begin{align}
	\alpha_{k} &=\alpha_{k|k-1} + \vert \boldsymbol{W}\vert,\\
	\beta_{k} &= \beta_{k|k-1} + 1.
	\end{align}
	
\subsection{Predicted likelihood for ET-PMB filter}
The predicted likelihood, used in the update step of the \ac{etpmb} filter (c.f. Sec.~\ref{sec:etpmbFilterUpdateStep} and Appendix~\ref{sec:Appendixetpmb}), for a set of detections $\boldsymbol{W}$ for a single \ac{et} is
\begin{align}
\ell_{\boldsymbol{W}} = \frac{\Gamma(\alpha_k)\beta_{k|k-1}^{\alpha_{k|k-1}}}{\Gamma(\alpha_{k|k-1})\beta_k^{\alpha_k}}
\prod_{l=1}^{\vert\boldsymbol{W}\vert} \mathcal{N}(\statez_{k,l} - \tilde{h}(\hat{\statey}_{k|k-1,l}),S_{k,l}),
\end{align}
where $\Gamma(\cdot)$ denotes the gamma function.

%% file: text/bios.tex

\begin{IEEEbiography}[{\includegraphics[width=1in,height=1.25in,clip,keepaspectratio]{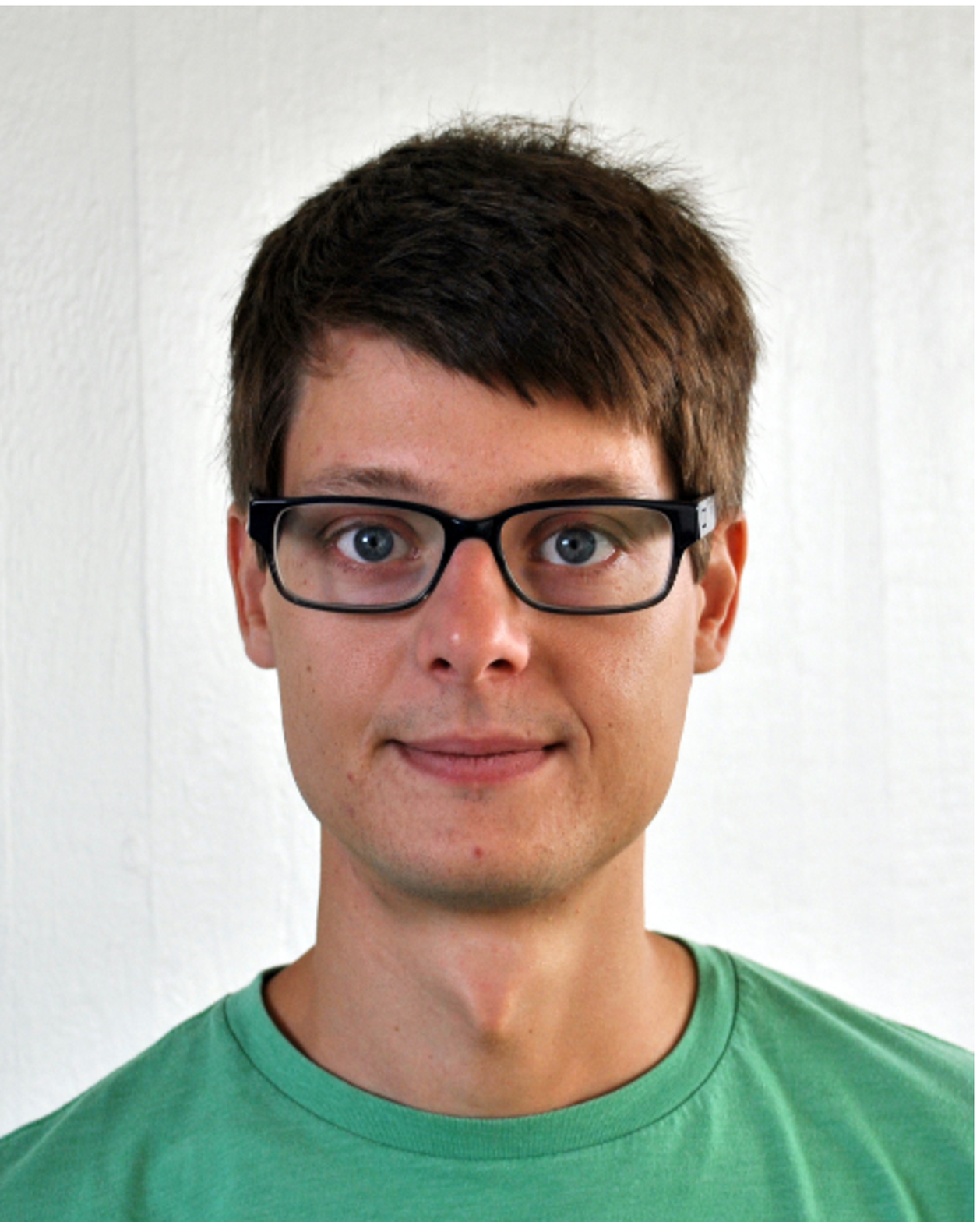}}]{Markus~Fr\"ohle}
	(S'11) received the B.Sc. and M.Sc. 
degrees in Telematics from Graz University of Technology, Graz, Austria, in 2009 and 2012, respectively.
He obtained the Ph.D. degree in Signals and Systems
from Chalmers University of Technology, Gothenburg, Sweden, in 2018. From 2012 to 2013, he was
with the Signal Processing and Speech Communication Laboratory, Graz University of Technology.
From 2013 to 2018, he was with the Department of
Electrical Engineering, Chalmers University of Technology. He joined Zenuity AB in 2019. His current
research interests include localization and tracking.
\end{IEEEbiography}
\begin{IEEEbiography}[{\includegraphics[width=1in,height=1.25in,clip,keepaspectratio]{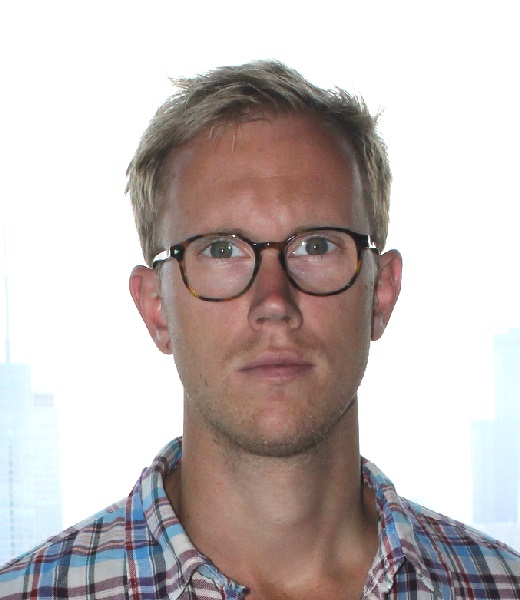}}]{Karl Granstr\"{o}m} (M'08) is a postdoctoral research fellow at the Department of Signals and Systems, Chalmers University of Technology, Gothenburg, Sweden. He received the MSc degree in Applied Physics and Electrical Engineering in May 2008, and the PhD degree in Automatic Control in November 2012, both from Link\"{o}ping University, Sweden. He previously held postdoctoral positions at the Department of Electrical and Computer Engineering at University of Connecticut, USA, from September 2014 to August 2015, and at the Department of Electrical Engineering of Link\"{o}ping University from December 2012 to August 2014. His research interests include estimation theory, multiple model estimation, sensor fusion and target tracking, especially for extended targets. He received paper awards at the Fusion 2011 and Fusion 2012 conferences. He has organised several workshops and tutorials on the topic Multiple Extended Target Tracking and Sensor Fusion. At Fusion 2018 the International Society for Information Fusion (ISIF) awarded him the ISIF Young Investigator Award for his contributions to extended target tracking research and his services to the research community.%
\end{IEEEbiography}

\begin{IEEEbiography}[{\includegraphics[width=1in,height=1.25in,clip,keepaspectratio]{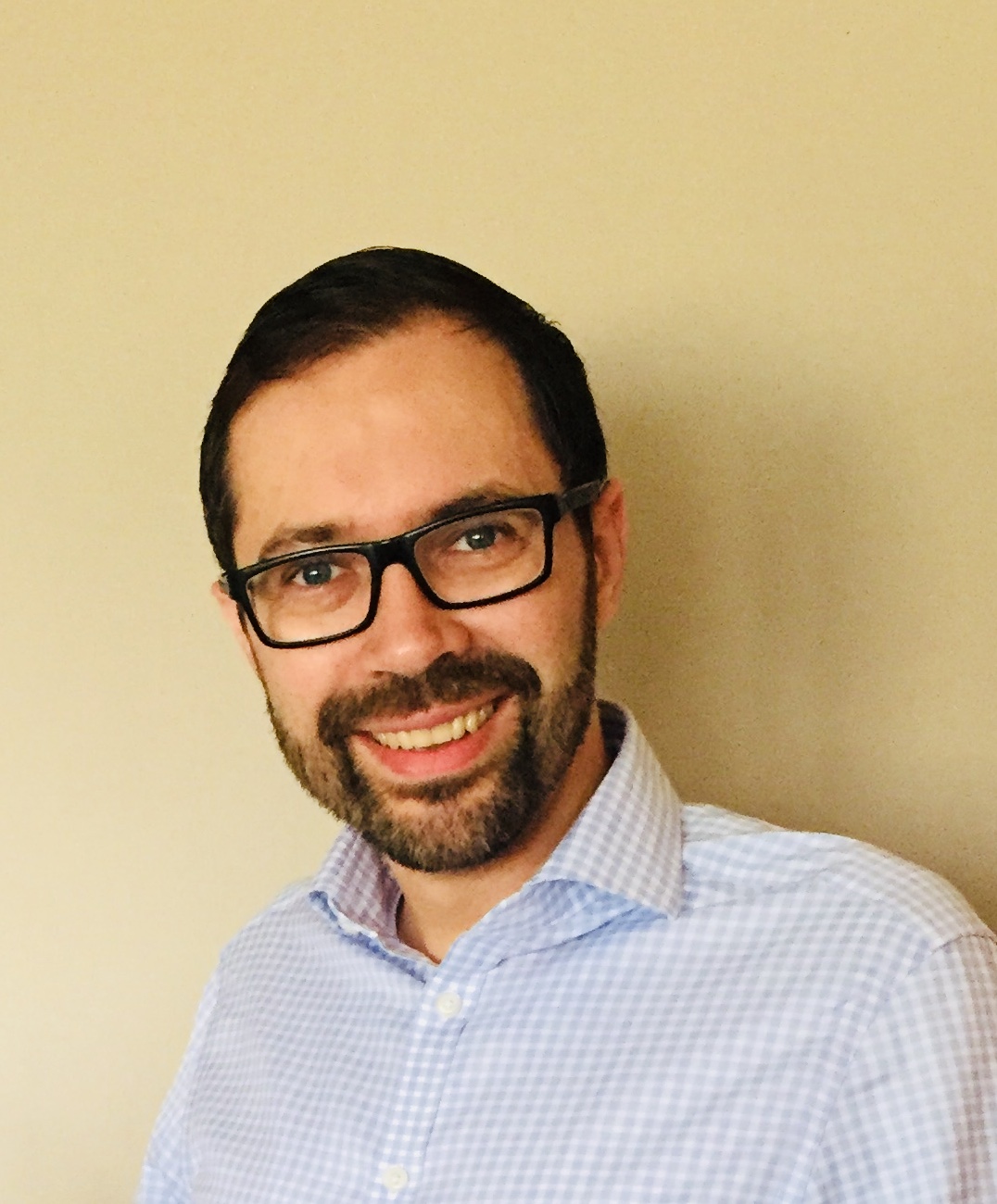}}]{Henk~Wymeersch}
	(S'01, M'05, SM'19) obtained the Ph.D. degree in Electrical Engineering/Applied Sciences in 2005 from Ghent University, Belgium. He is currently a Professor of Communication Systems with the Department of Electrical Engineering at Chalmers University of Technology, Sweden. Prior to joining Chalmers, he was a postdoctoral researcher from 2005 until 2009 with the Laboratory for Information and Decision Systems at the Massachusetts Institute of Technology. Prof. Wymeersch served as Associate Editor for IEEE Communication Letters (2009-2013), IEEE Transactions on Wireless Communications (since 2013), and IEEE Transactions on Communications (2016-2018). His current research interests include cooperative systems and intelligent transportation. 
\end{IEEEbiography}